 \documentclass{article}
\usepackage{bm,color}
\usepackage{hyperref}

\makeatletter
\def\oversortoftilde#1{\mathop{\vbox{\m@th\ialign{##\crcr\noalign{\kern3\p@}%
      \sortoftildefill\crcr\noalign{\kern3\p@\nointerlineskip}%
      $\hfil\displaystyle{#1}\hfil$\crcr}}}\limits}

\def\sortoftildefill{$\m@th \setbox\z@\hbox{$\braceld$}%
  \braceld\leaders\vrule \@height\ht\z@ \@depth\z@\hfill\braceru$}

\makeatother

\title{Second order nonlinear gyrokinetic theory : From the particle to the gyrocentre}

\author{Natalia Tronko$^{1,2}$, Cristel  Chandre$^{3}$,
\\
$^{1}$Max-Planck-Institut f\"{u}r Plasmaphysik,  85748 Garching, Germany,
\\
$^{2}$TU Munich, Mathematics Center, 85747, Garching, Germany,
\\
$^{3}$Aix Marseille Univ, CNRS, Centrale Marseille, I2M, Marseille, France}
\begin{document}
\maketitle 
\begin{abstract}
A gyrokinetic reduction is based on a specific ordering of the different small parameters characterizing the background magnetic field and the fluctuating electromagnetic fields. In this tutorial, we consider the following ordering of the small parameters: $\epsilon_B=\epsilon_\delta^2$ where $\epsilon_B$ is the small parameter associated with spatial inhomogeneities of the background magnetic field and $\epsilon_\delta$ characterizes the small amplitude of the fluctuating fields. In particular, we do not make any assumption on the amplitude of the background magnetic field. Given this choice of ordering, we describe a self-contained and systematic derivation which is particularly well suited for the gyrokinetic reduction, following a two-step procedure. We follow the approach developed in [Sugama, Physics of Plasmas 7, 466 (2000)]:
In a first step, using a translation in velocity, we embed the transformation performed on the symplectic part of the gyrocentre reduction in the guiding-centre one. In a second step, using a canonical Lie transform, we eliminate the gyroangle dependence from the Hamiltonian. As a consequence, we explicitly derive the fully electromagnetic gyrokinetic equations at the second order in $\epsilon_\delta$. 
\end{abstract}

\maketitle

\section{Introduction}
A strongly magnetized plasma forms a complex multi-scaled system in space and time. Building reduced models allows the identification of the main physical mechanisms in different regimes and configurations.

After more than three decades of active development, gyrokinetic theory is nowadays one of the important theoretical frameworks for the investigation of strongly magnetized plasmas. 
The main idea behind the gyrokinetic dynamical reduction consists of a systematic elimination of the smallest scales of motion, which leads to a drastic reduction of computational time.

From a theoretical viewpoint, the gyrokinetic reduction provides access to accurate predictions on long temporal and large spatial scale processes such as transport which is one of the main issues for fusion plasma confinement~\cite{Garbet_Idomura_2010}. 
In astrophysical plasmas, the gyrokinetic theory is also of interest~\cite{Schekochihin_2009}: Gyrokinetic simulations have been used to access small-scale spectra, in order to fill the gaps when magnetohydrodynamics approximations fail.

The gyrokinetic dynamical reduction exploits the fact that the particle dynamics is decomposed into a fast rotation around the magnetic field lines and a slow drift motion. The temporal scale of the gyromotion is set by the cyclotron frequency $\Omega=e B/m c$, where $e$ and $m$ are, respectively, the charge and mass of the particles, $B$ is the magnetic field amplitude and $c$ the speed of light.
The gyromotion is described by a fast gyroangle variable $\theta$ to which the slowly varying magnetic moment $\mu$ is canonically conjugate. At the lowest order,
\begin{equation}
\mu=\frac{m v_{\perp}^2}{2 B},
\end{equation}
where $v_{\perp}$ is the perpendicular velocity of the charged particle with respect to the magnetic field lines. The magnetic moment is a measure of the magnetic flux through the particle orbit. From this geometrical picture comes the idea of using $\mu$ as an action variable, canonically conjugated to the fast gyromotion around the magnetic field lines. In the case of a constant and uniform background magnetic field, $\mu$ is an exact dynamical invariant, regardless of the amplitude of the magnetic field. 

The sources for the violation of the conservation of the magnetic moment come from two different origins: first, spatial variations of the background quantities on the gyro-scales such as magnetic field non-uniformities and curvature and, second, the presence of electromagnetic fluctuations generated by the plasma. The goal of the gyrokinetic reduction is to reduce the dynamics taking into account these two sources of perturbations, and restore a conserved quantity, a modified magnetic moment. 

A fundamental aspect of the gyrokinetic theory is the assumption on the ordering of the various small parameters present in the system. To each choice of ordering will correspond a different reduced system and hence a different set of gyrokinetic equations to integrate numerically. The choice of ordering is driven by the specific geometry to be considered and by experimental observations. There are basically two groups of small parameters, each group associated with the two sources of perturbations which break the conservation of the magnetic moment: one group concerns the background magnetic field and another group characterizes the fluctuating fields. 

Roughly speaking, in the first group, $\epsilon_B$ is related to the background magnetic field inhomogeneities. The small parameter $\epsilon_B$ is given by $\epsilon_B=\rho_{\mathrm{th}}/L_B$, where $\rho_{\mathrm{th}}$ is the thermal Larmor radius and $L_B=\Vert\nabla B/B\Vert^{-1}$ defines the spatial scale at which the magnetic field exhibits significant variations. 

In the second group of parameters, $\epsilon_\delta$ is related to the amplitude of the fluctuating fields, $\epsilon_\parallel$ to the parallel gradients of these fields, and $\epsilon_\omega$ to their temporal variations: $\epsilon_{\delta}=(k_{\perp}\rho_{\mathrm{th}}) e\phi_1/T_i$, where $\phi_1$ is the amplitude of the fluctuating electrostatic potential, $k_\perp$ are perpendicular wavevectors of the fluctuation spectrum and $T_i$ the ion temperature. Furthermore, we assume that $\Vert {\bf A}_1\Vert$ is of the same order as $\phi_1 c/v_{\rm th}$ where $v_{\rm th}$ is a characteristic thermal velocity. The gyrokinetic theory assumes that $\epsilon_\delta$ is small.
In addition, experimental observations report that the most dangerous instabilities occur in the plane perpendicular to the background magnetic field for strongly magnetized plasmas. A small parameter takes this into account: $\epsilon_\parallel=E_{1\|}/E_{1\perp}\sim k_{\|}/ k_{\perp}$. A usual assumption is that the temporal variations of the fluctuating fields is small, so $\epsilon_\omega=\omega/\Omega$ is taken into account in the expansion, where $\omega$ is a characteristic frequency in the spectrum of the fluctuating fields and $\Omega$ is the ion cyclotron frequency. Other parameters for the fluctuating fields can be defined and we refer to~\cite{Brizard_Hahm} for a more complete discussion. 

The main assumption on the small parameters which sustain the derivation of the gyrokinetic model is an assumption which links both groups of parameters, the ones of the background magnetic field and the ones of the fluctuating fields, namely, $\epsilon_B\sim \epsilon_{\delta}^2$. This assumption is sustained by the majority of global gyrokinetic codes, e.g., ORB5~\cite{Jolliet_2007}, GKW~\cite{Peeters_2009}, GENE~\cite{Goerler_2011}, GYSELA~\cite{Grandgirard_2016} and GYRO~\cite{Candy_2003}. The main constraint behind this choice of ordering is to have simplified models suitable for their numerical implementation. It should be noted that recent experimental results for Tore Supra~\cite{Casati_2009} as well as numerical results obtained with ORB5 for large systems like ITER and DIII-D~\cite{Wersal_2012} indicate the necessity of retaining the maximal ordering $\epsilon_B\sim\epsilon_{\delta}$ for typical turbulence scales in the core of the plasma. The situation is different at the plasma edge, where the difference between both small parameters become more important, i.e., $\epsilon_{\delta}\sim 100\epsilon_{B}$, for typical edge turbulence scales.
 As a consequence, our reduction applies to systems where the magnetic field is not so strong, which is of interest, e.g., in astrophysical plasmas.
Concerning the parameters within the group of the fluctuating field parameters, we consider two usual assumptions: $\epsilon_\parallel\sim \epsilon_\delta$ and $\epsilon_\omega\sim \epsilon_\delta$.    

Given a particular choice of ordering, there exist several techniques to derive the gyrokinetic theory: direct ones and structural ones.
A direct approach akin to~\cite{CATTO1978,CATTO1981,Frieman_Chen_1982} performs an asymptotic expansion on the Vlasov equation, making the control of orderings and consistency rather cumbersome, for an alternative direct approach, see~\cite{Abel_2013}.
Structural derivations of gyrokinetic reductions follow Littlejohn's seminal work~ by using the Hamiltonian/Lagrangian framework. The implementation of the Hamiltonian formalism for the particle dynamical reduction towards the guiding-centre motion began with the works of Littlejohn in the Hamiltonian framework~\cite{Littlejohn_1979,Littlejohn_1981} and in the Lagrangian one~\cite{Littlejohn_1983}. The main advantage of these approaches is the consistency of the reduced particle model. An additional step towards gyrokinetic equations considers the coupling between the reduced particle dynamics with electromagnetic fields within the field-particle Lagrangian, providing a framework for the consistent derivation of the gyrokinetic models~\cite{Sugama_2000,brizard_prl_2000}; see also~\cite{Squire_Qin_2013,Burby_2015} for a Hamiltonian field formulation. Nowadays this framework is widely used for the derivation of consistently reduced models suitable for the numerical implementation~\cite{Brizard_Hahm,TBS_2016,PPCF_2017}.

The goal of the systematic gyrokinetic dynamical reduction consists in building a new set of phase-space variables, such that the $\theta$-dependence is completely decoupled from the other variables, and that the magnetic moment $\mu$ has trivial dynamics, i.e., $\dot{\mu}=0$. Therefore, the reduced particle dynamics is described in the $5$-dimensional phase space with variables $\left(\mathbf{X},v_\parallel,\mu\right)$ where $\mathbf{X}$ represents the reduced particle position (the gyrocentre), $v_\parallel$ is a variable which mirrors the parallel velocity of the particle and $\mu$ is the conserved magnetic moment. 
This change of coordinates is constructed via a perturbative series of phase-space transformations. The main advantage of this approach is that these transformations are invertible at each step of the perturbative procedure, allowing one to recover information on the particle dynamics from the averaged one. 

The phase-space Lagrangian is the starting point of the gyrokinetic derivation. It is given by
\begin{equation}
\label{eqn:L}
 L_0(\mathbf{x},\mathbf{v};t)=\left(\frac{e}{c}\mathbf{A}(\mathbf{x})+m\mathbf{v}\right)\cdot\dot{\mathbf x} -H(\mathbf{x},\mathbf{v},t), 
\end{equation}
where $\mathbf x$ and $\mathbf v$ are the position and velocity of the charged particle. The first term, proportional to $\dot{\bf x}$, represents the symplectic (or Liouville) part and the second term the Hamiltonian part.

Electromagnetic perturbations are introduced via the phase-space Lagrangian perturbation:
\begin{equation}
\label{eqn:L_pert}
L_{1}(\mathbf{x},\mathbf{v};t)= \epsilon_{\delta}\left(\frac{e}{c}\mathbf{A}_1(\mathbf{x},\epsilon_\delta t)\cdot\dot{\mathbf{x}}-e\phi_1(\mathbf{x},\epsilon_\delta t)\right),
\end{equation}
where we notice that the perturbation associated with the magnetic potential only affects the symplectic part of the phase-space Lagrangian and the electrostatic potential, its Hamiltonian part.

In \cite{Hahm_1988,Brizard_1989,Brizard_Hahm}, the standard gyrokinetic dynamical reduction is organized in two consecutive steps: the \textit{guiding-centre} reduction, where the new set of phase-space variables is constructed to evidence the conservation of the magnetic moment $\mu$ with respect to the background magnetic field nonlinearities, and the subsequent \textit{gyrocentre} reduction which builds a modified set of variables to restore the magnetic moment conservation broken by the introduction of the electromagnetic fluctuations.
This gyrokinetic dynamical reduction splits the difficulties in two steps with respect to the small parameter of the system: $\epsilon_B$ for the guiding-centre step and $\epsilon_{\delta}$ for the gyrocentre step.
Each of these steps consists in eliminating the gyrophase dependence from the symplectic and the Hamiltonian part, simultaneously.

In this tutorial, we follow an alternative two-step derivation of the gyrokinetic equations, as proposed in \cite{Grebogi_1979,Sugama_2000},
which allows treating the symplectic and the Hamiltonian part of the Lagrangian in two consecutive steps.
This is achieved by an apt shift of the velocity, followed by a modified guiding-centre reduction to move the $\theta$-dependence of the symplectic part of the Lagrangian to the Hamiltonian part. Subsequently, the $\theta$-dependence of the Hamiltonian is removed by some canonical Lie transforms. This approach has also been the one followed in~\cite{PPCF_2017} in the particular case where the electromagnetic potential does not have any perpendicular component. This method is very well suited for our choice of ordering.
 
 This tutorial is organised as follows: In Sec.~\ref{sec:gyrocentre} we introduce the velocity shift which allows the application of the guiding-centre theory following \cite{Littlejohn_1983}. In Sec.~\ref{sec:Lie-transform} we recall the general procedure of canonical Lie transforms and we apply this perturbative procedure to derive the reduced Hamiltonian dynamics at the second order in $\epsilon_\delta$, explicitly providing the new set of variables at order $\epsilon_\delta$.  
 Finally, in Sec.~\ref{sec:Maxwell-Vlasov}, we present the gyrocentre characteristics, from which the gyrokinetic Vlasov equation is reconstructed, and we briefly remind the variational method to derive the reduced Maxwell equations. For completeness, we add an appendix containing a slightly revisited guiding-centre derivation, with the full details of the derivation, which is suited for our choice of ordering. As a result, the derivation of the gyrokinetic model proposed in this tutorial is self-consistent, and does not rely on previous knowledge of the guiding-centre theory.

\section{Gyrocentre as a modified guiding-centre}
\label{sec:gyrocentre}

The perturbed one-form associated with the perturbed Lagrangian~(\ref{eqn:L})-(\ref{eqn:L_pert}) is given by
\begin{equation}
\gamma_{\rm pert} = \left(\frac{e}{c}{\bf A}({\bf x})+\epsilon_\delta \frac{e}{c}{\bf A}_1({\bf x},\epsilon_\delta t)+m{\bf v}\right)\cdot {\rm d}{\bf x}-H {\rm d}t,
\end{equation}
where 
\begin{equation}
\label{eq:H}
H=\frac{1}{2}m{\bf v}^2+\epsilon_\delta e\phi_1({\bf x},\epsilon_\delta t).
\end{equation}
The part in ${\rm d}t$ is referred to as the Hamiltonian part of the one-form, whereas the part in ${\rm d}{\bf x}$ is its symplectic (Liouville) part. 
We translate the particle velocity ${\bf v}$:
\begin{equation}
\bar{\bf v}={\bf v}+\epsilon_\delta\frac{e}{mc}{\bf A}_1({\bf x},\epsilon_\delta t).
\end{equation}
This velocity shift \cite{Grebogi_1979,Sugama_2000} allows us to apply directly and readily Littlejohn's guiding-centre theory~\cite{Littlejohn_1983} without performing the reduction calculations twice (first, the guiding-centre reduction without the perturbation ${\bf A}_1$ and $\phi_1$, and then the gyrocentre reduction) in the symplectic part of the one-form. 
Here we can now apply the guiding-centre results on $\gamma_{\rm pert}$ in the variables $({\bf x},\bar{\bf v})$. We decompose $\bar{\bf v}$ into
\begin{equation}
\bar{\bf v}=\bar{v}_\parallel \hat{\bf b}({\bf x})+\bar{v}_\perp \hat{\bm \perp}(\bar{\theta},{\bf x}),
\end{equation}
where $\hat{\bf b}={\bf B}/B$. The above-relation defines the two vectors $\hat{\bm\perp}$ and $\hat{\bm\rho}$ of the orthonormal basis $(\hat{\bm\perp},\hat{\bm\rho},\hat{\bf b})$. Given an orthonormal basis $(\hat{\bf b}_1({\bf x}),\hat{\bf b}_2({\bf x}),\hat{\bf b}({\bf x}))$, the following vectors  
\begin{eqnarray}
&& \hat{\bm \perp}(\bar{\theta}, {\bf x})= -\hat{\bf b}_1({\bf x})\sin \bar{\theta} -\hat{\bf b}_2({\bf x})\cos \bar{\theta},\label{eq:defperp}\\
&& \hat{\bm \rho}(\bar{\theta}, {\bf x})= \hat{\bf b}_1({\bf x})\cos \bar{\theta} -\hat{\bf b}_2({\bf x})\sin \bar{\theta},\label{eq:defrho}
\end{eqnarray}
define the angle $\bar{\theta}$. 
The crucial step in the guiding-centre theory is a shift of the particle position, i.e., ${\bf x}=\bar{\bf X}+ {\bm\rho}$ with 
\begin{equation}
\label{eqn:rhoshift}
{\bm \rho}= \frac{m\bar{v}_\perp c}{eB(\bar{\bf X})}\hat{\bm\rho}(\bar{\theta},\bar{\bf X})+  \bar{\bm \rho}_1+{\cal O}(\epsilon_B^2),
\end{equation}
where the explicit expression of $\bar{\bm \rho}_1$ which is of order $\epsilon_B$ is given in Appendix [for more details on the guiding-centre theory, we refer to~\cite{Littlejohn_1983,Cary_Brizard,Tronko_Brizard_2015}]. Here the expression for $\bar{\bm \rho}_1$ is not explicitly needed since we only provide the change of coordinates at order $\epsilon_\delta$ (see Sec.~\ref{sec:Lie-transform}). 
After the modified guiding-centre reduction, the one-form becomes
\begin{equation}
\bar{\gamma}_{\rm pert} = \left(\frac{e}{c}{\bf A}(\bar{\bf X})+m\bar{v}_\parallel \hat{\bf b}(\bar{\bf X})-\frac{mc}{e}\mu {\bf R}^*\right)\cdot {\rm d}\bar{\bf X}+\frac{mc}{e}\bar{\mu}{\rm d}\bar{\theta}-H {\rm d}t +{\cal O}(\epsilon_B^2),
\end{equation}
where $\bar{\mu}=m\bar{v}_\perp^2/(2B(\bar{\bf X}))$, and ${\bf R}^*=\nabla \hat{\bf b}_1\cdot \hat{\bf b}_2+(\hat{\bf b}\cdot \nabla \times \hat{\bf b})\hat{\bf b}/2$. We notice that the symplectic part of $\bar{\gamma}_{\rm pert}$ has no explicit dependence on $\bar{\theta}$, which was the objective of the guiding-centre reduction. The Poisson bracket associated with the symplectic part of $ \bar{\gamma}_{\rm pert}$  is given by
\begin{equation}
\label{eq:bracGC}
\{F,G\}_{\rm gc}=\frac{e}{mc}\left( \frac{\partial F}{\partial \bar{\theta}}\frac{\partial G}{\partial \bar{\mu}}-\frac{\partial F}{\partial \bar{\mu}}\frac{\partial G}{\partial \bar{\theta}}\right)+\frac{{\bf B}^*}{mB_\parallel^*}\cdot \left( \nabla^* F\frac{\partial G}{\partial \bar{v}_\parallel}-\frac{\partial F}{\partial \bar{v}_\parallel}\nabla^* G\right)-\frac{c\hat{\bf b}}{eB_\parallel^*}\cdot (\nabla^* F\times \nabla^* G),
\end{equation}
for observables $F$ and $G$, functions of $(\bar{\bf X},\bar{\theta},\bar{\mu}, \bar{v}_\parallel)$, and where 
\begin{eqnarray}
&&\nabla^*=\nabla-\mathbf{R}^*\ \frac{\partial}{\partial \theta},\\
&&{\bf B}^*={\bf B}+\frac{mc}{e}\bar{v}_\parallel\nabla\times \hat{\bf b}-\frac{mc^2}{e^2}\bar{\mu} \nabla\times {\bf R}^*,
\end{eqnarray}
and $B_\parallel^*=\hat{\bf b}\cdot {\bf B}^*$. 
After the translation in velocity, Hamiltonian~(\ref{eq:H}) becomes
\begin{equation}
\label{eq:Hnew}
H=\frac{1}{2}m\bar{v}_\parallel^2+\bar{\mu}B(\bar{\bf X})+\epsilon_\delta e\psi_1(\bar{\bf X},\bar{\theta},\bar{\mu},\bar{v}_\parallel,t)+\epsilon_\delta^2  \frac{e^2}{2mc^2}\Vert {\bf A}_1(\bar{\bf X}+{\bm \rho},\epsilon_\delta t)\Vert^2,
\end{equation}
where there is an explicit $\theta$-dependence at order $\epsilon_\delta$ and $\epsilon_\delta^2$ through the potentials $\psi_1$ and ${\bf A}_1$.
The modified potential $\psi_1$ is given by
\begin{eqnarray}
\psi_1(\bar{\bf X},\bar{\theta},\bar{\mu},\bar{v}_\parallel,t)&=&\phi_1(\bar{\bf X}+{\bm \rho},\epsilon_\delta t) -\frac{\bar{v}_\parallel}{c}\hat{\bf b}(\bar{\bf X}+{\bm \rho})\cdot {\bf A}_1(\bar{\bf X}+{\bm \rho},\epsilon_\delta t)\nonumber \\
&& \qquad -\sqrt{\frac{2\bar{\mu}B(\bar{\bf X})}{mc^2}}\hat{\bm \perp}(\bar{\theta},\bar{\bf X}+{\bm \rho})\cdot {\bf A}_1(\bar{\bf X}+{\bm \rho},\epsilon_\delta t). 
\end{eqnarray}
It is important to note that all the fluctuating part has been removed from the symplectic part of the one-form and moved to the Hamiltonian part. In this way, the averaging over the fast variable $\bar{\theta}$ has to be performed only on the Hamiltonian and not on the symplectic part of the one-form. By using canonical transformations, the symplectic part of the one-form is not affected (up to an exact one-form). We notice that Hamiltonian~(\ref{eq:Hnew}) has an explicit time-dependence, through the fluctuating potentials $\phi_1$ and ${\bf A}_1$. Therefore in order to perform canonical transformations, it is more convenient to autonomize the system, i.e., by considering that $t$ is an additional dynamical variable and introducing $k$ its canonically conjugate variable. The extended Hamiltonian becomes   
\begin{equation}
{\cal H}=\frac{1}{2}m\bar{v}_\parallel^2+\bar{\mu}B(\bar{\bf X})+\epsilon_\delta e\psi_1(\bar{\bf X},\bar{\theta},\bar{\mu},\bar{v}_\parallel,t)+\epsilon_\delta^2  \frac{e^2}{2mc^2}\Vert {\bf A}_1(\bar{\bf X}+{\bm \rho},\epsilon_\delta t)\Vert^2 +k,
\end{equation}
and the extended Poisson bracket becomes
\begin{equation}
\label{eq:bracGCt}
\{{\cal F},{\cal G}\}=\{{\cal F},{\cal G}\}_{\rm gc}+\frac{\partial {\cal F}}{\partial t}\frac{\partial {\cal G}}{\partial k}-\frac{\partial {\cal F}}{\partial k}\frac{\partial {\cal G}}{\partial t},
\end{equation}
where $\{\cdot,\cdot \}_{\rm gc}$ is given by Eq.~(\ref{eq:bracGC}). In what follows, the observables in the extended phase space, i.e., functions of $(\bar{\bf X},\bar{\theta},\bar{\mu},\bar{v}_\parallel,t,k)$, are denoted with a calligraphic lettering, whereas functions of $(\bar{\bf X},\bar{\theta},\bar{\mu},\bar{v}_\parallel,t)$ will be denoted in roman lettering. 

As a final remark, we notice that the shift in the position [see Eq.~(\ref{eqn:rhoshift})] contains the Larmor radius from the guiding centre and also the displacement generated by the perturbation field ${\bf A}_1$, and explicitly depends on ${\bf A}_1$. At the leading order we have:
\begin{equation}
\label{eq:xpart}
{\bf x}=\bar{\bf X}+\frac{mc}{eB}\hat{\bf b}\times \bar{\bf v}=\bar{\bf X}+ \frac{mc}{eB}\hat{\bf b}\times {\bf v}+\epsilon_\delta \frac{1}{B}\hat{\bf b}\times {\bf A}_1+ {\cal O}(\epsilon_\delta^2).
\end{equation}
Furthermore, the averaging procedure performed in the Hamiltonian will modify the position $\bar{\bf X}$ into the position of the gyrocenter ${\bf X}_{\rm gy}$. We will come back to the expressions of the new coordinates after performing the averaging procedure. 

\section{Averaging procedure of the Hamiltonian}
\label{sec:Lie-transform}

\subsection{Canonical Lie transforms}

In order to perform the averaging with respect to the fast variable $\bar{\theta}$ in the Hamiltonian, we use canonical Lie transforms. These transforms are near-identity canonical changes of coordinates which do not modify the expression of the symplectic part of the one-form (up to an exact one form), or equivalently, do not change the expression of the Poisson bracket. A canonical Lie transform only affects the Hamiltonian, and with an apt choice of generating function eliminates the unwanted part of the Hamiltonian, in our case, its fast-varying part. For more details on Lie transforms, we refer to~\cite{Cary_1981}. 
The invertible change of coordinates from the old variables ${\bf Z}=(\bar{\bf X},\bar{\theta},\bar{\mu},\bar{v}_\parallel,t,k)$ to the new (gyrokinetic) ones ${\bf Z}_{\rm gy}=({\bf X}_{\rm gy},\theta_{\rm gy},v_{\parallel {\rm gy}},t,k_{\rm gy})$ is defined as
\begin{equation}
{\bf Z}_{\rm gy} = {\rm e}^{ \pounds_{S}}{\bf Z},
\end{equation}
where $S$ is the scalar generating function of the transformation which is chosen as a function of $(\bar{\bf X},\bar{\theta},\bar{\mu},\bar{v}_\parallel,t) $, and the operator $\pounds_S$ is defined as $\pounds_{S}=\{ S,\cdot \}$. The bracket $\{\cdot,\cdot\}$ is given by Eqs.~(\ref{eq:bracGC})-(\ref{eq:bracGCt}). 
This change of coordinates transforms any observable ${\cal F}({\bf Z})$ into $\bar{\cal F}({\bf Z}_{\rm gy})$ according to
\begin{equation}
\label{eq:Fbar}
\bar{\cal F}({\bf Z}_{\rm gy})={\rm e}^{-\pounds_{S}}{\cal F}({\bf Z}_{\rm gy})={\cal F}-\{ S,{\cal F} \}+\frac{1}{2}\{S,\{ S,{\cal F} \} \}+\mathcal{O}(S^3), 
\end{equation}
which is obtained from the scalar invariance $\bar{{\cal F}}({\bf Z}_{\rm gy})={\cal F}({\bf Z})$ (and the fact that the Poisson bracket satisfies the Leibniz rule) and where the right-hand-side of Eq.~(\ref{eq:Fbar}) is evaluated at ${\bf Z}_{\rm gy}$. 

We recall that the guiding-centre Poisson bracket~(\ref{eq:bracGC}) is decomposed into
\begin{equation}
\{F,G\}_{\rm gc}= \{F,G\}_{-1}+\{F,G\}_{0} + \{F,G\}_{1},
\end{equation} 
and the Hamiltonian $H=H_0+\epsilon_\delta H_1+\epsilon_\delta^2 H_2$ [see Eq.~(\ref{eq:Hnew})]. In order to remove the $\bar{\theta}$-dependence from $H$, we consider a generating function of the type $S=\epsilon_\delta  S_1+\epsilon_\delta^2 S_2$. The purpose of $S_1$ is to eliminate the fluctuating part of the Hamiltonian $H$ at order $\epsilon_\delta$ (i.e., present in $H_1$) and $ S_2$ eliminates the fluctuating terms at order $\epsilon_\delta^2$. In order to illustrate the method, we first consider the order $\epsilon_\delta$. We decompose $H_1$ in a fluctuating and an averaged part:  $H_1=\widetilde{H_1}+ \langle H_1\rangle$, where $\langle H_1\rangle= (2\pi)^{-1}\int_0^{2\pi} {\rm d}\bar{\theta} H_1$.
At order $\epsilon_\delta$, it leads to
\begin{equation}
\bar{H}=H-\{S,H\}_{\rm gc}-\frac{\partial S}{\partial t}+O(S^2),
\end{equation}
i.e.,
\begin{equation}
\bar{H}=H_0+\epsilon_\delta\left( \langle H_1\rangle +\widetilde{H_1}-\left\{S_1, H_0\right\}_{-1}-\left\{S_1, H_0\right\}_0-\left\{S_1, H_0\right\}_1-\frac{\partial S_1}{\partial t}\right)+\mathcal{O}(\epsilon_{\delta}^2). \label{Lie_transform_1}
\end{equation}
By inspecting the various terms in the above-equation, we notice that $\{S_1, H_0\}_1$ is of order $\epsilon_B$ since it involves $\nabla B$. This term is thus neglected even at the next order since, according to our ordering, $\epsilon_B\sim \epsilon_\delta^2$.  
The term $\left\{S_1, H_0\right\}_0$ involves a term proportional to ${\bf B}^*\cdot \nabla S_1$. Up to order $\epsilon_B$, this term is the parallel gradient of the generating function. Since the generating function is a function of the fluctuating fields, this term will be of order of the parallel gradients of the fluctuating fields $\phi_1$ and ${\bf A}_1$, which are assumed to be of order $\epsilon_\delta$, so the term $\left\{S_1, H_0\right\}_0$ is moved to order $\epsilon_\delta^2$. In addition, we assume that $\partial S_1/\partial t$ is of order $\epsilon_\delta$ which comes from the ordering $\epsilon_\omega\sim\epsilon_\delta$. Therefore the resulting equation which determines the generating function $S_1$ is 
\begin{equation}
\label{eqn:S11}
\{S_1,H_0\}_{-1}=\frac{eB}{mc}\frac{\partial S_1}{\partial \bar{\theta}}=\widetilde{H_1}. 
\end{equation}
At the leading order, the new Hamiltonian becomes
\begin{equation}
\bar{H}=H_0+\epsilon_\delta \langle H_1\rangle+O(\epsilon_\delta^2).
\end{equation}
We extend this analysis to the second order, where the expansion of the Hamiltonian is given by Eq.~(\ref{eq:Fbar}) as
\begin{equation}
\label{eq:HH2}
\bar{H}= H-\{S,H\}_{\rm gc}-\frac{\partial S}{\partial t}+\frac{1}{2}\{S,\{S,H\}_{\rm gc}\}_{\rm gc}+\frac{1}{2}\left\{S,\frac{\partial S}{\partial t}\right\}_{\rm gc}+O(S^3).
\end{equation}
The terms containing $\partial S/\partial t$ comes from the fact that the transformation is time-dependent [and is taken care by the extended bracket (\ref{eq:bracGCt})]. The term $\{S,\partial S/\partial t\}_{\rm gc}$ in Eq.~(\ref{eq:HH2}) is neglected since it is of order $\epsilon_\delta^3$ given that $\epsilon_\omega\sim \epsilon_\delta$. 
The expansion of the Hamiltonian at the order $\epsilon_\delta^2$ becomes
\begin{eqnarray}
\bar{H}&=&H_0+\epsilon_\delta \langle H_1\rangle-\epsilon_\delta\left(\frac{\partial S_1}{\partial t}+ \left\{S_1, H_0\right\}_0 \right)
+\epsilon_\delta^2\left(H_2-\frac{\partial S_2}{\partial t}\right. \nonumber \\
&& \left. -\{S_1,H_1\}_{\rm gc}-\{S_2,H_0\}_{\rm gc}+\frac{1}{2}\{S_1,\{S_1,H_0\}_{\rm gc}\}_{\rm gc} \right)+O(\epsilon_\delta^3).
\end{eqnarray}
We eliminate the term $\epsilon_\delta^2 \partial S_2/\partial t$ since it is of order $\epsilon_{\delta}^3$ given that $\epsilon_\omega\sim\epsilon_\delta$. In the term $\{S_2,H_0\}_{\rm gc}$, only the term $\{S_2,H_0\}_{-1}$ matters since $\{S_2,H_0\}_{0}$ and $\{S_2,H_0\}_{1}$ are higher order using the same argument as above for $S_1$. Using the same argument, in the term $\{S_1,\{S_1,H_0\}_{\rm gc}\}_{\rm gc}$, only the terms $\{S_1,\{S_1,H_0\}_{-1}\}_{-1}$ and $\{S_1,\{S_1,H_0\}_{-1}\}_{1}$ remain, since the term $\{S_1,\{S_1,H_0\}_{-1}\}_{0}=\{S_1,\widetilde{H_1}\}_0$ involves parallel gradients of $S_1$ or $H_1$ (of order $\epsilon_\delta$). This leads to
\begin{eqnarray}
&&\bar{H}=H_0+\epsilon_\delta \langle H_1\rangle -\epsilon_\delta\left(\frac{\partial S_1}{\partial t}+ \left\{S_1, H_0\right\}_0 \right)+\epsilon_\delta^2\left(\langle H_2\rangle +\widetilde{H_2}-\left\{S_2, H_0\right\}_{-1} -\{S_1,H_1\}_{-1}\right.\nonumber \\
&& \qquad \quad \left.- \{S_1,H_1\}_1
+\frac{1}{2}\left\{S_1,\left\{S_1, H_0\right\}_{-1}\right\}_{-1} +\frac{1}{2}\left\{S_1,\left\{S_1, H_0\right\}_{-1}\right\}_{1} \right)+\mathcal{O}(\epsilon_{\delta}^3).
\end{eqnarray}
 We choose $S_2$ such that it eliminates the fluctuating part of $\bar{H}$ at order $\epsilon_\delta^2$. 
The equation which determines $S_2$ is then
\begin{eqnarray}
\left\{S_2, H_0\right\}_{-1}&=&\widetilde{H_2}-\{S_1,\langle H_1 \rangle\}_{-1}-\frac{1}{2}\oversortoftilde{\{S_1,\widetilde{H_1}\}_{-1}}-\epsilon_\delta^{-1}\left( \frac{\partial S_1}{\partial t}+\{S_1,H_0\}_0 \right)\nonumber \\
&& \qquad -\{S_1,\langle H_1\rangle\}_1-\frac{1}{2}\oversortoftilde{\{S_1,\widetilde{H_1}\}_1},
\end{eqnarray}
where we have used Eq.~(\ref{eqn:S11}).
Consequently, the new Hamiltonian becomes
\begin{equation}
\bar{H}=H_0+\epsilon_\delta \langle H_1 \rangle+\epsilon_\delta^2 \left( \langle H_2 \rangle -\frac{1}{2}\langle \{S_1, \widetilde{H_1}\}_{-1} \rangle -\frac{1}{2}\langle \{S_1,\widetilde{H_1}\}_1\rangle\right)+\mathcal{O}(\epsilon_{\delta}^3).
\end{equation}

\subsection{Application to Hamiltonian~(\ref{eq:Hnew})}
We rewrite the Poisson bracket $\{S_1,\widetilde{H_1}\}_{-1}$ as
\begin{equation}
\{S_1,\widetilde{H_1}\}_{-1}=\frac{e}{mc}\frac{\partial}{\partial \bar{\mu}}\left(\widetilde{H_1}\frac{\partial S_1}{\partial \bar{\theta}} \right)- \frac{e}{mc}\frac{\partial}{\partial \bar{\theta}}\left(\widetilde{H_1}\frac{\partial S_1}{\partial \bar{\mu}} \right).
\end{equation}
From the expression of $S_1$, we conclude that the averaged Hamiltonian $\bar{H}$ obtained from $H=H_0+\epsilon_\delta H_1+\epsilon_\delta^2 H_2$ is
\begin{equation}
\bar{H}=H_0+\epsilon_\delta \langle H_1 \rangle+\epsilon_\delta^2 \left(\langle H_2 \rangle -\frac{1}{2B}\frac{\partial}{\partial \bar{\mu}} \left\langle \widetilde{H_1}^2\right\rangle +\frac{c}{2eB}\hat{\bf b}\cdot \langle \nabla S_1\times \nabla \widetilde{H_1}\rangle\right) +\mathcal{O}(\epsilon_{\delta}^3),
\end{equation}
where we have used Eq.~(\ref{eqn:S11}) and where we notice the presence of two additional second order terms, compared to the na\"{i}ve average of the Hamiltonian.
Next, we apply this result to Hamiltonian~(\ref{eq:Hnew}) where, in the old coordinates $\bf Z$,
\begin{eqnarray}
&& H_0 = \bar{\mu}B(\bar{\bf X})+\frac{1}{2}m \bar{v}_{\parallel}^2,\\
&& H_1 = e\psi_1(\bar{\bf X},\bar{\theta},\bar{\mu},\bar{v}_{\parallel},t),\\
&& H_2 = \frac{e^2}{2mc^2}\Vert {\bf A}_1(\bar{\bf X}+{\bm \rho},\epsilon_\delta t)\Vert^2.
\end{eqnarray} 
In the new coordinates ${\bf Z}_{\rm gy}$, the reduced Hamiltonian is then
\begin{eqnarray}
H_{\rm gy}&=& H_{\rm gc}+\epsilon_\delta e\langle \psi_1 \rangle \nonumber \\
&& + \epsilon_\delta^2 \left(\frac{e^2}{2mc^2}\langle \Vert{\bf A}_1({\bf X}_{\rm gy}+{\bm \rho}_{\rm gy},\epsilon_\delta t)\Vert^2\rangle-\frac{e^2}{2B({\bf X}_{\rm gy})}\frac{\partial}{\partial \mu_{\rm gy}} \left\langle \widetilde{\psi_1}^2\right\rangle\right. \nonumber \\
&& \qquad\qquad  -\left.\frac{mc^2}{2B({\bf X}_{\rm gy})^2}\hat{\bf b}({\bf X}_{\rm gy})\cdot \left\langle \nabla \widetilde{\psi_1}\times \int {\rm d}\theta_{\rm gy}\nabla \widetilde{\psi_1}\right\rangle \right), \label{eqn:Hgyy}
\end{eqnarray}
where 
\begin{equation}
\label{eqn:rhogy}
{\bm\rho}_{\rm gy}=\frac{mc}{eB({\bf X}_{\rm gy})}\sqrt{\frac{2{\mu_{\rm gy}}B({\bf X}_{\rm gy})}{m}}\hat{\bm\rho}({\theta}_{\rm gy},{\bf X}_{\rm gy}).
\end{equation} 
In Eq.~(\ref{eqn:Hgyy}), the function $\psi_1$ is evaluated at $({\bf X}_{\rm gy},{\theta}_{\rm gy},{\mu}_{\rm gy},{v}_{\parallel \rm gy},t)$, and $H_{\rm gc}$ is given by
\begin{equation}
H_{\rm gc}=\mu_{\rm gy}B({\bf X}_{\rm gy})+\frac{1}{2}mv_{\parallel {\rm gy}}^2.
\end{equation}
The averaging has been performed using the generating function $S_1$ given by
\begin{equation}
S_1({\bf X}_{\rm gy},{\theta}_{\rm gy},{\mu}_{\rm gy},{v}_{\parallel \rm gy},t)=\frac{mc}{B}\int {\rm d}{\theta}_{\rm gy}\,  \widetilde{\psi}_1({\bf X}_{\rm gy},{\theta}_{\rm gy},{\mu}_{\rm gy},{v}_{\parallel \rm gy},t),
\end{equation}
which is essential in order to determine, at the leading order, the change of coordinates which has realized the reduction. 

We recover the expression of the Hamiltonian obtained in \cite{Brizard_1989,Sugama_2000,Brizard_Hahm}. 

\subsection{Changes of coordinates}

Next, we look at the expression of the change of coordinates which links the particle dynamics with the gyrocentre dynamics. We recall that, at the leading order, this change is a result of two steps: a translation of the velocity by the perturbation fields and of the position by a modified Larmor radius, and an averaging performed at the Hamiltonian level using a canonical Lie transform. Below, we provide the explicit expressions at order $\epsilon_\delta$. 

Given our choice of generating function and the ordering of the bracket, the old coordinates ${\bf Z}$ as functions of the new (gyrokinetic) ones ${\bf Z}_{\rm gy}$ are written as
\begin{equation}
{\bf Z}={\bf Z}_{\rm gy}- \epsilon_\delta \{S_1,{\bf Z}_{\rm gy}\}_{\rm gc}+\mathcal{O}(\epsilon_{\delta}^2).
\end{equation}
We remind that there was a first step (a modified guiding-centre step) which mapped $({\bf x},{\bf v})$ into $(\bar{\bf X},\bar{\theta},\bar{\mu},\bar{v}_\parallel)$:
\begin{eqnarray}
&& {\bf x} = \bar{\bf X}+\frac{mc}{eB(\bar{\bf X})}\sqrt{\frac{2\bar{\mu}B(\bar{\bf X})}{m}}\hat{\bm\rho}(\bar{\theta},\bar{\bf X}) , \\
&& {\bf v} = \bar{v}_\parallel \hat{\bf b}(\bar{\bf X})+ \sqrt{\frac{2\bar{\mu}B(\bar{\bf X})}{m}}\hat{\bm \perp}(\bar{\theta},\bar{\bf X})-\epsilon_\delta \frac{e}{mc}{\bf A}_1(\bar{\bf X}+ {\bm\rho},\epsilon_\delta t).\label{eq:vexp}
\end{eqnarray}
The second step performed the averaging of the Hamiltonian using a canonical Lie transform. It mapped $(\bar{\bf X},\bar{\theta},\bar{\mu},\bar{v}_\parallel)$ into $({\bf X}_{\rm gy},{\theta}_{\rm gy},{\mu}_{\rm gy},{v}_{\parallel {\rm gy}})$. Up to order ${\cal O}(\epsilon_\delta^2)$, the expressions for this second change of coordinates are
\begin{eqnarray}
&&{\bf X}_{\rm gy}= \bar{\bf X}+\epsilon_\delta \{S_1,\bar{\bf X}\}_0+\epsilon_\delta\{S_1,\bar{\bf X}\}_1, \nonumber\\
&& \qquad  =\bar{\bf X}-\epsilon_\delta \left( \frac{1}{m}\hat{\bf b}\frac{\partial S_1}{\partial \bar{v}_\parallel} + \frac{c}{eB}\hat{\bf b}\times \nabla S_1\right),\\ 
&&{\bf \theta}_{\rm gy}= \bar{\theta}+\epsilon_\delta \{S_1,\bar{\theta}\}_{-1}=\bar{\theta}-\epsilon_\delta\frac{e}{mc}\frac{\partial S_1}{\partial \bar{\mu}},\\
&& {\mu}_{\rm gy} = \bar{\mu}+\epsilon_\delta \{S_1,\bar{\mu}\}_{-1}=\bar{\mu}+\epsilon_\delta\frac{e}{B}\widetilde{\psi}_1,\\
&&{v}_{\parallel {\rm gy}} = \bar{v}_\parallel. 
\end{eqnarray}
By combining the two changes of coordinates, we obtain
\begin{eqnarray}
{\bf x} &=& {\bf X}_{\rm gy}+{\bm\rho}_{\rm gy}-\epsilon_\delta \{S_1,{\bf X}_{\rm gy}+{\bm\rho}_{\rm gy} \}_{\rm gc} \nonumber \\
&=& {\bf X}_{\rm gy}+ {\bm\rho}_{\rm gy}+\epsilon_\delta\left[ -\frac{1}{B}\hat{\bf b}\int {\rm d}\theta_{\rm gy} \widetilde{A_{1\parallel}} +\frac{c}{eB}\hat{\bf b}\times \nabla S_1 \right. \nonumber \\
&& \qquad \qquad \left.-\frac{1}{B}\sqrt{\frac{mc^2}{2\mu_{\rm gy}B}} \left(\widetilde{\psi_1}\hat{\bm\rho}-2\mu_{\rm gy}\hat{\bm\perp}\int {\rm d}\theta_{\rm gy} \frac{\partial \widetilde{\psi_1}}{\partial \mu_{\rm gy}} \right) \right],\\
{\bf v} &=& {v}_{\parallel {\rm gy}} \hat{\bf b}-\frac{eB}{mc}\hat{\bf b}\times{\bm\rho}_{\rm gy}+\epsilon_\delta \frac{eB}{mc}\hat{\bf b}\times \{S_1,{\bm\rho}_{\rm gy}\}_{-1} -\epsilon_\delta \frac{e}{mc}{\bf A}_{1},
\end{eqnarray}
where the functions $\hat{\bf b}$, $B$, $\hat{\bm\rho}$ and $\hat{\bm\perp}$ are taken at ${\bf X}_{\rm gy}$, and the functions $\widetilde{\psi}_1$, $S_1$ and its derivatives, at $({\bf X}_{\rm gy},{\theta}_{\rm gy},{\mu}_{\rm gy},{v}_{\parallel {\rm gy}})$. The fields ${\bf A}_1$ and $\phi_1$ are evaluated at ${\bf X}_{\rm gy}+ {\bm\rho}_{\rm gy}$, where ${\bm\rho}_{\rm gy}$ is given by Eq.~(\ref{eqn:rhogy}).

We introduce the variables $({\theta},{\mu},{v}_{\parallel})$ associated with the velocity $\bf v$:
\begin{equation}
{\bf v}=v_\parallel \hat{\bf b}({\bf x})+\sqrt{\frac{2\mu B({\bf x})}{m}}\hat{\bm \perp}(\theta,{\bf x}).
\end{equation}
The expressions of the variables $({\theta},{\mu},{v}_{\parallel})$ are given by
\begin{eqnarray}
&& \theta = \theta_{\rm gy}+\epsilon_\delta \frac{e}{mc}\left(\frac{\partial S_1}{\partial \mu_{\rm gy}}+\sqrt{\frac{m}{2\mu_{\rm gy}B}}\hat{\bm\rho}\cdot{\bf A}_1 \right),\\
&& \mu=\mu_{\rm gy}-\epsilon_\delta \frac{e}{B}\left(\widetilde{\psi_1}+\frac{1}{c}\sqrt{\frac{2\mu_{\rm gy}B}{m}}\hat{\bm\perp}\cdot{\bf A}_1\right),\\
&& v_\parallel=v_{\parallel {\rm gy}}-\epsilon_\delta \frac{e}{mc} {A}_{1\parallel}. 
\end{eqnarray}
From these expressions, we deduce the relation between ${\bm \rho}_{\rm gy}$ and the Larmor radius:
\begin{equation}
\frac{mc}{eB}\hat{\bf b}\times {\bf v}={\bm \rho}_{\rm gy}-\epsilon_\delta\{S_1,{\bm \rho}_{\rm gy}\}_{-1}-\epsilon_\delta \frac{1}{B}\hat{\bf b}\times {\bf A}_1.
\end{equation} 

In the two-step reduction procedure we presented, the complexity of the derivation is shared between the Hamiltonian and the symplectic parts of the phase-space Lagrangian: First, the velocity shift allows one to move all the gyroangle dependencies from the symplectic part to the
Hamiltonian, so as to apply readily the guiding center transform. Then a series of canonical Lie transforms are applied to the Hamiltonian
to finalize the dynamical gyrocenter reduction. An alternative derivation uses a general reduction method applied to the Hamiltonian and the symplectic part at the same time, which requires applying non-canonical Lie transforms to the differential forms (see e.g. \cite{Brizard_Hahm}).

\section{Gyrokinetic Vlasov-Maxwell equations}
\label{sec:Maxwell-Vlasov}
We derive the gyrokinetic Maxwell-Vlasov equations following \cite{Sugama_2000} which provides the following Lagrangian:
\begin{equation}
\mathcal{L}= \sum_{\mathrm{sp}}\int {\rm d}V_0\  {\rm d}W_0\ F({\mathbf Z}_0,t_0) L_p(\mathbf{Z}_{\rm gy}(\mathbf{Z}_0,t_0;t),\dot{\mathbf{Z}}_{\rm gy}(\mathbf{Z}_0,t_0;t),t) +\int{\rm d} V \frac{|{\mathbf E}|^2-|{\mathbf B}+ \epsilon_\delta\nabla\times{\bf A}_1|^2}{8 \pi},
\label{field-gyrocenter_L}
\end{equation}
where
\begin{equation}
 L_{\rm p}=\left(\frac{e}{c}\mathbf{A}+m v_{\| {\rm gy}}\hat{\mathbf b} \right)\cdot\dot{\mathbf{X}}_{\rm gy}+ \frac{mc}{e}\mu_{\rm gy}\dot{\theta}_{\rm gy}-H_{\rm gy},
\end{equation}
and where the gyrocentre distribution function of the species $\rm{sp}$ $F(\mathbf{Z}_0,t_0)$ is defined at the arbitrary initial gyrocentre phase-space position $\mathbf{Z}_0$ and arbitrary initial time $t_0$. We will not use a specific notation to distinguish the distribution functions of the different species for simplicity.
The reduced phase-space variables are ${\mathbf Z}_{\rm gy}=({\mathbf X}_{\rm gy},v_{\|{\rm gy}},\mu_{\rm gy})$ and the phase-space volume element is given by ${\rm d}\Omega = {\rm d}V_0 {\rm d} W_0$ with ${\rm d}V_0$ denoting the volume element in physical space, i.e., ${\rm d}V_0= {\rm d}^3{\mathbf X}_{\rm gy}$ for the gyrocentre part and ${\rm d}V={\rm d}^3{\mathbf x}$ for the electromagnetic part; the velocity gyrocentre phase-space volume is ${\rm d}W_0= B_{\|}^*(\mathbf{Z}_0) {\rm d} v_{\| {\rm gy}} {\rm d}\mu_{\rm gy} {\rm d}\theta_{\rm gy}$.

We perform the change of variables $\mathbf{Z}_{\rm gy}=\mathbf{Z}_{\rm gy}(\mathbf{Z}_0,t_0;t)$, such that the first term of Eq.~(\ref{field-gyrocenter_L}) becomes:
$$
\sum_{\mathrm{sp}}\int {\rm d}\Omega \ F({\mathbf Z}_{\rm gy},t) L_{\rm p}(\mathbf{Z}_{\rm gy},\dot{\mathbf{Z}}_{\rm gy},t).
$$
The gyrokinetic Vlasov equation is obtained using the gyrocentre characteristics, from the conservation of the distribution function along the trajectories, i.e., 
\begin{equation}
\frac{{\rm d}}{{\rm d} t} F(\mathbf{Z}_{\rm gy}(\mathbf{Z}_0,t_0;t),t)=\frac{\partial}{\partial t} F(\mathbf{Z}_{\rm gy},t)+\dot{{\mathbf Z}}_{\rm gy}\cdot\frac{\partial}{\partial {\mathbf Z}_{\rm gy}} F(\mathbf{Z}_{\rm gy},t)=0.
\label{eqn:Vlasov_cons}
\end{equation}
For the electromagnetic part of the Lagrangian, we use the quasi-neutrality and Darwin approximation [see, e.g.,~\cite{Bottino_Sonnendruecker, Krause_2007} for more details] which boils down to neglecting the $E^2$ term in the Lagrangian. The resulting expression for the gyrokinetic Lagrangian used for the derivation of the gyrokinetic Poisson and Amp\`ere equations is:
\begin{equation}
{\mathcal L}=\sum_{\mathrm{sp}}\int {\rm d}\Omega\ F({\mathbf Z}_{\rm gy},t) L_{\rm p}(\mathbf{Z}_{\rm gy},\dot{\mathbf{Z}}_{\rm gy},t)
-\int{\rm d} V\ \frac{|\nabla\times\left({\mathbf A}_0+\epsilon_{\delta}{\mathbf A}_1\right)|^2}{8\pi}+O(\epsilon_\delta^3).
\label{GK_action}
\end{equation}
%
\subsection{Gyrokinetic Vlasov-Maxwell equations}
In this section, we provide the gyrokinetic Vlasov-Maxwell equations in the weak form, since this form is well suited to the finite-element discretization, necessary for PIC Monte-Carlo simulations as performed in ORB5.

The gyrokinetic quasineutrality equation is obtained in the weak form using the functional derivatives of the action:
\begin{eqnarray}
0&=&\frac{\delta\mathcal L }{\delta\phi_1}\circ\widehat{\phi}_1 = 
\epsilon_\delta \sum_{\mathrm{sp}} \int {\rm d}\Omega\ F  \left(  -e\langle\widehat{\phi}_1\rangle
+ \epsilon_{\delta}\frac{e^2}{B}\frac{\partial}{\partial \mu_{\rm gy}}\left( \langle \Psi_1 \widehat{\phi}_1\rangle - \langle \Psi_1 \rangle \langle \widehat{\phi}_1\rangle\right)\right. \nonumber \\
&& \left.\qquad +  \epsilon_\delta \frac{mc^2}{2B^2}\hat{\bf b}\cdot \left\langle \nabla \widetilde{\widehat{\phi}_1}\times \int {\rm d}\theta_{\rm gy} \nabla \widetilde{\Psi_1}+ \nabla \widetilde{\Psi_1}\times \int {\rm d}\theta_{\rm gy} \nabla \widetilde{\widehat{\phi}_1}\right\rangle \right),
\label{eq:Poisson} 
\end{eqnarray}
where $\widehat{\phi}_1$ is a test function, evaluated at ${\bf X}_{\rm gy}+{\bm \rho}_{\rm gy}$. 

The parallel component of the gyrokinetic Amp\`ere equation is given by
\begin{eqnarray}
0&=&\frac{\partial\mathcal L}{\partial A_{1\|}}\circ{\widehat{A}_{1\|}}
=-\frac{\epsilon_{\delta}}{4\pi}\int {\rm d}V\ \widehat{A}_{1\|}\ \widehat{\bf b}\cdot \nabla\times{\mathbf B}+
\epsilon_\delta \sum_{\mathrm{sp}} e \int {\rm d}\Omega\ F \frac{v_{\|{\rm gy}}}{c}\langle \widehat{A}_{1\|}\rangle
\nonumber
\\
&-&\frac{\epsilon_{\delta}^2}{4\pi}\int {\rm d}V\ \nabla \widehat{A}_{1\|}\cdot
[\hat{\bf b}\times (\nabla\times {\bf A}_{1})] -\epsilon_\delta^2 \sum_{\mathrm{sp}} \frac{e^2}{m c^2}\int {\rm d}\Omega\ F  \langle A_{1\|}\widehat{A}_{1\|}\rangle \nonumber
\\
&-&
\epsilon_\delta^2  \sum_{\mathrm{sp}} \int {\rm d}\Omega\ F \frac{v_{\|{\rm gy}}}{c}  \frac{e^2}{B}\frac{\partial}{\partial \mu_{\rm gy}}\left( \langle \Psi_1 \widehat{A}_{1\|}\rangle -\langle \Psi_1\rangle \langle\widehat{A}_{1\|}\rangle\right)\label{eq:Ampere1} \\
&-& \epsilon_\delta^2 \sum_{\mathrm{sp}} \int {\rm d}\Omega\ F \frac{mc^2}{2B^2}\frac{v_{\|{\rm gy}}}{c}\hat{\bf b}\cdot \left\langle \nabla \widetilde{\widehat{A}_{1\|}}\times \int {\rm d}\theta_{\rm gy} \nabla \widetilde{\Psi_1}+ \nabla \widetilde{\Psi_1}\times \int {\rm d}\theta_{\rm gy} \nabla \widetilde{\widehat{A}_{1\|}}\right\rangle, 
\nonumber
\end{eqnarray}
where $\widehat{A}_{1\parallel}$ is a test function, evaluated at ${\bf X}_{\rm gy}+{\bm \rho}_{\rm gy}$ when the integral is over ${\rm d}\Omega$, and evaluated at ${\bf X}_{\rm gy}$ when the integral is over ${\rm d}V$.
The perpendicular component of the gyrokinetic Amp\`ere equation is given by:
\begin{eqnarray}
0&=&\frac{\delta\mathcal L}{\delta \mathbf{A}_{1\perp}}\circ{\widehat{\mathbf{A}}_{1\perp}}
\nonumber
= 
-\frac{\epsilon_{\delta}}{4\pi}\int {\rm d}V\ {\mathbf A}_{1\perp}\cdot (\nabla\times\mathbf{B})
+\epsilon_\delta \sum_{\mathrm{sp}} e \int {\rm d}\Omega \ F \ \sqrt{\frac{2\mu_{\rm gy}B}{mc^2}}\langle \hat{\bm\perp}\cdot \widehat{\mathbf A}_{1\perp}\rangle 
\\
&-&\frac{\epsilon_{\delta}^2}{4\pi}\int {\rm d}V\ (\nabla\times\mathbf{A}_{1})\cdot (\nabla\times\widehat{\mathbf A}_{1\perp})-\epsilon_\delta^2\sum_{\mathrm{sp}}\frac{e^2}{mc^2} \int {\rm d}\Omega\ F \langle {\bf A}_{1}\cdot \widehat{\mathbf A}_{1\perp} \rangle  \nonumber
\\
&-& \epsilon_\delta^2 \sum_{\mathrm{sp}} e \int {\rm d}\Omega \ F \frac{e}{B}\frac{\partial}{\partial \mu_{\rm gy}}\left( \sqrt{\frac{2\mu_{\rm gy}B}{mc^2}} 
(\langle \Psi_1 \hat{\bm\perp}\cdot \widehat{\mathbf A}_{1\perp}\rangle -\langle \Psi_1\rangle \langle \hat{\bm\perp}\cdot \widehat{\mathbf A}_{1\perp}\rangle) \right) \nonumber \\
&&  -\epsilon_\delta^2\sum_{\mathrm{sp}} \int {\rm d}\Omega\ F \frac{mc^2}{2B^2}\sqrt{\frac{2\mu_{\rm gy}B}{mc^2}}\hat{\bf b}\cdot \left\langle \nabla (\oversortoftilde{\hat{\bm\perp}\cdot \widehat{\mathbf A}_{1\perp}})\times \int {\rm d}\theta_{\rm gy} \nabla \widetilde{\Psi_1} \right. \label{eq:Ampere2}\\
&&\nonumber \qquad \qquad \qquad \qquad \qquad \qquad \qquad \qquad \left. + \nabla \widetilde{\Psi_1}\times \int {\rm d}\theta_{\rm gy} \nabla (\oversortoftilde{\hat{\bm\perp}\cdot \widehat{\mathbf A}_{1\perp}})\right\rangle,
\end{eqnarray}
where $\widehat{\mathbf A}_{1\perp}$ is a test function, evaluated at ${\bf X}_{\rm gy}+{\bm \rho}_{\rm gy}$ when the integral is over ${\rm d}\Omega$, and evaluated at ${\bf X}_{\rm gy}$ when the integral is over ${\rm d}V$.

We remark that both Amp\'ere's laws (\ref{eq:Ampere1}) and (\ref{eq:Ampere2}) are identical to the results presented in \cite{Sugama_2000} and \cite{Brizard_Hahm}.

\subsection{\label{sec:GK_characteristics}Gyrokinetic particle characteristics and the gyrokinetic Vlasov equation}
The gyrokinetic Vlasov equation is obtained from the gyrocentre particle characteristics following Eq.~(\ref{eqn:Vlasov_cons}) and taking into account that ${\rm d}{\mathbf Z}/{\rm d}t=\{\mathbf{Z},H_{\rm gy}\}_{\rm{gc}}$.
The gyrocenter characteristics are given by:
\begin{eqnarray}
\label{char_X}
\dot{\mathbf X}_{\rm gy}=\left\{{\mathbf{X}_{\rm gy}},{H}_{\rm gy}\right\}_{\mathrm{gc}}&=& \frac{c\widehat{\mathbf b}}
{e B_{\|}^{*}}\times\nabla {H}_{\rm gy}+\frac{\mathbf{B}^{*}}{mB_{\|}^{*}}\ \frac{\partial {H}_{\rm gy}}{\partial{{v}_{\| {\rm gy}}}},\\
\label{char_p}
\dot{v}_{\| {\rm gy}}=\left\{v_{\| {\rm gy}},{H}_{\rm gy}\right\}_{\mathrm{gc}}&=&-\frac{\mathbf{B}^{*}}{mB_{\|}^{*}}\cdot \nabla {H}_{\rm gy},
\end{eqnarray}
where $H_{\rm gy}$ is given by Eq.~(\ref{eqn:Hgyy}). 
The fully nonlinear gyrokinetic Vlasov equation for the Vlasov distribution $F({\bf X}_{\rm gy},v_{\parallel {\rm gy}},\mu_{\rm gy},t)$ is therefore given by:
\begin{eqnarray}
0=\frac{{\rm d} F}{{\rm d}t}&=&\frac{\partial F}{\partial t}+\{{\mathbf X}_{\rm gy},H_{\rm gy}\}_{\mathrm{gc}}\cdot\nabla F + \{v_{\| {\rm gy}},H_{\rm gy}\}_{\mathrm{gc}}\ \frac{\partial F}{\partial v_{\| {\rm gy}}},
\end{eqnarray}
where we notice that the term in $\partial F/\partial \mu_{\rm gy}$ is absent since $\{\mu_{\rm gy},H_{\rm gy}\}_{\rm gc}=0$.
The Vlasov equation can be rewritten with the help of the Poisson bracket~(\ref{eq:bracGC}):
\begin{equation}
\label{Vlasov_bracket}
\frac{\partial F}{\partial t}=-\{F,H_{\rm gy}\}_{\mathrm{gc}}.
\end{equation}

Equation~(\ref{Vlasov_bracket}) for the Vlasov equation and Eqs.~(\ref{eq:Poisson}), (\ref{eq:Ampere1}) and (\ref{eq:Ampere2}) for the Maxwell equations constitute the second-order gyrokinetic Vlasov-Maxwell equations associated with the second order Hamiltonian for the gyrocentres given by Eq.~(\ref{eqn:Hgyy}), consistent with the ordering $\epsilon_B\sim \epsilon_\delta^2$.  
We notice that due to the quasi-neutrality and the Darwin approximations, the gyrokinetic Poisson and  Amp\`ere equations do not contain explicit time derivatives.

\section*{Acknowledgments}
The authors would like to thank  C.~Angioni, A.~Bottino, A.~J.~Brizard, J.~W.~Burby, A.~Mishchenko, B.~D.~Scott and E.~Sonnendr\"ucker for helpful discussions. 
This work has been carried out within the framework of
the EUROfusion Consortium and has received funding from
the Euratom research and training Programme No. 2014-
2018 under Grant Agreement No. 633053. The views and
opinions expressed herein do not necessarily reflect those of
the European Commission.

\appendix

\section{Littlejohn's guiding-centre theory}
\label{Littlejohn_theory}
In this Appendix, we revisit Littlejohn's guiding-centre theory, following~\cite{Littlejohn_1983}. We begin with the one-form for the motion of the particle:
\begin{equation}
 \gamma=\left(\frac{e}{c}{\bf A}({\bf x})+m{\bf v}\right)\cdot {\rm d}{\bf x}-H {\rm d}t.
\end{equation}
The revisit comes from the fact that we do not consider $\epsilon$ as a small parameter as it was done in \cite{Littlejohn_1979,Littlejohn_1981,Littlejohn_1983}. Here the small parameter is $\epsilon_B$ which relates to the spatial variation of the external magnetic field.  

The idea of the guiding centre is to find a change of coordinates which removes the fluctuating part from the one-form. The guiding-centre transformation is based on two different changes of coordinates, a far-from-identity transformation which consists of a shift of the position by the Larmor radius, and a near-identity transformation which eliminates the fluctuating terms of the one-form at order $\epsilon_B$. 

The first step is a translation by the Larmor radius, inspired from the situation where the magnetic field is constant and uniform: 
\begin{equation}
{\bf x}=\bar{\bf x}+ {\bm \rho}_0.
\end{equation}
We translate the velocity, by defining
\begin{equation}
{\bf w}={\bf v}+\frac{e}{mc} \left[{\bf A}(\bar{\bf x}+ {\bm \rho}_0)-{\bf A}(\bar{\bf x})- ({\bm \rho}_0\cdot \nabla ) {\bf A}(\bar{\bf x}) -\frac{1}{2}({\bm\rho}_0{\bm\rho}_0:\nabla \nabla){\bf A}(\bar{\bf x}) \right],
\end{equation}
where the last term written in indices is $-(1/2) \rho_i\rho_j \partial^2 {\bf A}/\partial \bar{x}_i\partial \bar{x}_j $.
We notice that the quantity with which the velocity has been translated is of order $\epsilon_B^2$ (since it involves third derivatives of the vector potential, i.e., second derivatives of the magnetic field). 
We decompose the velocity ${\bf w}$ in the following way:
\begin{equation}
{\bf w}=w_\parallel \hat{\bf b}({\bf x})+ { w}_\perp \hat{\bm\perp}(\theta,{\bf x}),
\end{equation}
where the orthonormal basis $(\hat{\bf b},\hat{\bm\perp},\hat{\bm\rho})$ is defined in Sec.~\ref{sec:gyrocentre}. 
We decompose the one-form as 
\begin{equation}
\gamma = \gamma_0+{\gamma}_1, 
\end{equation}
where 
\begin{equation}
 \gamma_0= \left(\frac{e}{c}{\bf A}(\bar{\bf x})+m w_\parallel\hat{\bf b}(\bar{\bf x})\right)\cdot {\rm d}\bar{\bf x}-H {\rm d}t,
\end{equation}
and
\begin{eqnarray}
 {\gamma}_1&=& \left( \frac{e}{c}({\bm \rho}_0\cdot \nabla ) {\bf A}+\frac{e}{2 c}({\bm\rho}_0{\bm\rho}_0:\nabla \nabla){\bf A}+ m{w}_\perp \hat{\bm\perp}(\theta,\bar{\bf x}+ {\bm \rho}_0) +m w_\parallel [\hat{\bf b}(\bar{\bf x}+ {\bm \rho}_0)-\hat{\bf b}(\bar{\bf x})] \right)\cdot {\rm d}\bar{\bf x} \nonumber \\
&& +\left( \frac{e}{c}{\bf A} + \frac{e}{c}({\bm \rho}_0\cdot \nabla ) {\bf A}+\frac{e}{2 c}({\bm\rho}_0{\bm\rho}_0:\nabla \nabla){\bf A} + m{w}_\perp \hat{\bm\perp}(\theta,\bar{\bf x}+ {\bm \rho}_0) +m w_\parallel\hat{\bf b}(\bar{\bf x}+ {\bm \rho}_0)\right) \cdot {\rm d}{\bm \rho}_0.\nonumber \\
&& 
\end{eqnarray} 
We use a gauge-invariance to simplify the one form: ${\gamma}$ can be replaced by ${\gamma}+{\rm d}\sigma$ where $\sigma$ is any scalar function of $(\bar{\bf x},w_\parallel, {\bf w}_\perp)$. Looking at the shape of ${\gamma}_1$, especially the terms in ${\rm d}{\bm \rho}_0$, some terms are removed by considering
\begin{equation}
\label{eqn:S}
 \sigma_1=-\frac{e}{c}{\bf A}\cdot {\bm \rho}_0- \frac{e}{2c}({\bm \rho}_0\cdot \nabla){\bf A} \cdot {\bm \rho}_0-\frac{e}{6 c}({\bm\rho}_0{\bm\rho}_0:\nabla \nabla){\bf A}\cdot {\bm \rho}_0. 
\end{equation}
In what follows and otherwise specified, the dependence of the functions on the variables is omitted when these functions are unambiguously expressed in the current set of variables. 
The one-form becomes
\begin{eqnarray}
 {\gamma}_1+{\rm d}\sigma_1&=&\left( \frac{eB}{c}\hat{\bf b}\times {\bm \rho}_0+\frac{e}{2 c} ({\bm\rho}_0\cdot \nabla) (B\hat{\bf b})\times {\bm\rho}_0 +m{w}_\perp \hat{\bm\perp}+mw_\perp ({\bm\rho}_0\cdot \nabla )\hat{\bm\perp} \right. \nonumber \\
&& \qquad \left.+ mw_\parallel ({\bm \rho}_0\cdot \nabla)\hat{\bf b}  \right)\cdot {\rm d}\bar{\bf x}\nonumber \\
&& +\left(\frac{eB}{2c}\hat{\bf b}\times {\bm \rho}_0+\frac{e}{3c} ({\bm\rho}_0\cdot \nabla) (B\hat{\bf b})\times {\bm\rho}_0+m{w}_\perp \hat{\bm\perp}+mw_\perp ({\bm\rho}_0\cdot \nabla )\hat{\bm\perp} +mw_\parallel \hat{\bf b}\right.\nonumber\\
&& \qquad+ \left. mw_\parallel ({\bm \rho}_0\cdot \nabla)\hat{\bf b}\right)\cdot {\rm d}{\bm\rho}_0,
\end{eqnarray}
where we have neglected the contributions of order $\epsilon_B^2$. In the above expression we have used the identity
\begin{equation}
({\bm \rho}_0\cdot \nabla) {\bf A}- \nabla {\bf A} \cdot {\bm\rho}_0=B\hat{\bf b}\times {\bm \rho}_0. 
\end{equation}
The condition for ${\bm \rho}_0$ is 
\begin{equation}
 \frac{eB}{c}\hat{\bf b}\times {\bm \rho}_0+m{w}_\perp\hat{\bm\perp}=0,
\end{equation}
i.e., 
\begin{equation}
 {\bm\rho}_0= \frac{m w_\perp c}{eB}\hat{\bm\rho}.
\end{equation}
The one-form becomes 
\begin{eqnarray}
{\gamma}_1+{\rm d}\sigma_1&=& \frac{m^2c}{eB}\left(\frac{w_\perp^2}{2B} (\hat{\bm\rho}\cdot \nabla) (B\hat{\bf b})\times \hat{\bm\rho}+w_\perp^2(\hat{\bm\rho}\cdot\nabla)\hat{\bm\perp}+ w_\parallel w_\perp (\hat{\bm \rho}\cdot \nabla)\hat{\bf b} \right)\cdot {\rm d}\bar{\bf x}\nonumber \\
&& +  \frac{m^2c}{eB}\left[w_\parallel \hat{\bf b}+\frac{1}{2}{w}_\perp\hat{\bm\perp}+ \frac{mc}{eB}\left( \frac{w_\perp^2}{3B} (\hat{\bm\rho}\cdot \nabla) (B\hat{\bf b})\times \hat{\bm\rho}+w_\perp^2(\hat{\bm\rho}\cdot\nabla)\hat{\bm\perp}\right.\right.\nonumber 
\\
&&\qquad\qquad+ \left.\left. w_\parallel w_\perp (\hat{\bm \rho}\cdot \nabla)\hat{\bf b}\right)  \right] \cdot {\rm d}(w_\perp\hat{\bm\rho}).
\end{eqnarray}
The leading order term (in the small parameter $\epsilon_B$) is 
\begin{equation}
 \frac{m^2c}{eB}\left( w_\parallel \hat{\bf b}+\frac{1}{2}{ w}_\perp\hat{\bm\perp} \right)\cdot {\rm d}( w_\perp \hat{\bm\rho}) = \frac{m^2c}{eB}\left[\frac{w_\perp^2}{2} {\rm d}\theta -\left( \frac{w_\perp^2}{2}  \nabla \hat{\bm\perp}\cdot \hat{\bm\rho}-w_\perp w_\parallel \nabla \hat{\bm\rho}\cdot \hat{\bf b}\right) \cdot {\rm d}\bar{\bf x}\right],
\end{equation}
since $\hat{\bm\perp}=\partial\hat{\bm\rho}/\partial\theta$ and where we have used $ \nabla \hat{\bm\rho}\cdot \hat{\bm\perp}= -\nabla \hat{\bm\perp}\cdot \hat{\bm\rho}$. 
We notice that the terms in $ {\rm d}\bar{\bf x}$ are of order $\epsilon_B$. We rewrite the one-form $\gamma $ as 
\begin{equation}
\gamma+{\rm d}\sigma_1= \tilde{\gamma}_0+ \tilde{\gamma}_1, 
\end{equation}
where 
\begin{equation}
 \tilde{\gamma}_0= \left(\frac{e}{c}{\bf A}(\bar{\bf x})+m w_\parallel\hat{\bf b}(\bar{\bf x})\right)\cdot {\rm d}\bar{\bf x}+\frac{m^2w_\perp^2 c}{2eB}  {\rm d}\theta -H {\rm d}t,
\end{equation}
is the leading order and
\begin{eqnarray}
\tilde{\gamma}_1&&=  \frac{m^2c}{eB}\left(\frac{w_\perp^2}{2B} (\hat{\bm\rho}\cdot \nabla) (B\hat{\bf b})\times \hat{\bm\rho}+w_\perp^2(\hat{\bm\rho}\cdot\nabla)\hat{\bm\perp}-\frac{w_\perp^2}{2}  \nabla \hat{\bm\perp}\cdot \hat{\bm\rho}+ w_\parallel w_\perp (\nabla\times \hat{\bf b})\times \hat{\bm\rho} \right)\cdot {\rm d}\bar{\bf x}\nonumber \\
&& +\frac{m^3c^2}{e^2B^2}\left( \frac{w_\perp^2}{3B} (\hat{\bm\rho}\cdot \nabla) (B\hat{\bf b})\times \hat{\bm\rho} +w_\perp^2(\hat{\bm\rho}\cdot\nabla)\hat{\bm\perp}\right.\nonumber\\
&&\qquad\qquad \qquad\qquad\qquad\left.+ w_\parallel w_\perp (\hat{\bm \rho}\cdot \nabla)\hat{\bf b} \right) \cdot (\hat{\bm\rho}{\rm d}w_\perp +w_\perp \hat{\bm\perp} {\rm d}\theta ), \label{eqn:gamma1}
\end{eqnarray}
is of order $\epsilon_B$. Here we have used the identity $(\hat{\bm\rho}\cdot \nabla)\hat{\bf b}+\nabla\hat{\bm\rho}\cdot \hat{\bf b}=(\nabla \times \hat{\bf b})\times \hat{\bm \rho}$. We notice that the order $\epsilon_B$ of the one-form, i.e., $\tilde{\gamma}_1$, contains fluctuating terms in $\theta$. These terms are eliminated by a near-identity transformation which we consider at order one in $\epsilon_B$:
\label{eq:cccc}
\begin{eqnarray}
&& \bar{\bf x}={\bf X}+ {\bm \xi}({\bf X},W_\parallel,W_\perp,\Theta),\label{eq:cX}\\
&& w_\parallel = W_\parallel + {\cal W}_\parallel ({\bf X},W_\parallel,W_\perp,\Theta),\label{eq:cP}\\
&&  w_\perp =  W_\perp + {\cal W}_\perp ({\bf X},W_\parallel, W_\perp,\Theta),\label{eq:cp}\\
&& \theta = \Theta+ {\cal T} ({\bf X},W_\parallel, W_\perp,\Theta),\label{eq:ct}
\end{eqnarray}
where the unknown functions ${\bm \xi}$, ${\cal W}_\parallel$, ${\cal W}_\perp$ and ${\cal T}$ are of order $\epsilon_B$. 

Since $\tilde{\gamma}_1$ is of order $\epsilon_B$, its expression in the new coordinates $({\bf X},W_\parallel,W_\perp,\Theta)$ is exactly the same as in the old coordinates $(\bar{\bf x},w_\parallel,w_\perp,\theta)$ up to $\epsilon_B^2$ terms. It remains to expand $\tilde{\gamma}_0$ at the first order in $\epsilon_B$. The expansion leads to 
\begin{equation}
\tilde{\gamma}_0=\left(\frac{e}{c}{\bf A}({\bf X})+m W_\parallel\hat{\bf b}({\bf X})\right)\cdot {\rm d}{\bf X}+ \frac{m^2 W_\perp^2 c}{2eB} {\rm d}\Theta -H {\rm d}t+\tilde{\gamma}_2,
\end{equation}
where 
\begin{eqnarray}
\tilde{\gamma}_2 &=&  \left(\frac{e}{c}({\bm \xi}\cdot \nabla) {\bf A}+m{\cal W}_\parallel \hat{\bf b} \right)\cdot {\rm d}{\bf X}+\left(\frac{e}{c}{\bf A}+m W_\parallel\hat{\bf b}\right)\cdot {\rm d}{\bm \xi}\nonumber \\
&& +\frac{m^2 c}{eB}W_\perp {\cal W}_\perp  {\rm d}\Theta +\frac{m^2W_\perp^2 c}{2eB}{\rm d}{\cal T}. 
\end{eqnarray}
We notice that we have used the approximation $B({\bf X}+{\bm\xi})\approx B({\bf X})$ since the gradients of $B$ are of order $\epsilon_B$ and ${\bm\xi}$ is also of order $\epsilon_B$, so the difference $B({\bf X}+{\bm\xi})- B({\bf X})$ is of order $\epsilon_B^2$. The same approximation holds for $\hat{\bf b}$. 
Using a gauge transformation similar to Eq.~(\ref{eqn:S}), we simplify $\tilde{\gamma}_2$ into
\begin{eqnarray}
\tilde{\gamma}_2+{\rm d}\sigma_2&=& \left(\frac{eB}{c}\hat{\bf b}\times{\bm \xi}+m{\cal W}_\parallel \hat{\bf b} \right)\cdot {\rm d}{\bf X}+m W_\parallel\hat{\bf b}\cdot {\rm d}{\bm \xi}\nonumber \\
&& +\frac{m^2 c}{eB}W_\perp {\cal W}_\perp  {\rm d}\Theta +\frac{m^2W_\perp^2 c}{2eB}{\rm d}{\cal T}, \label{eqn:gamma2}
\end{eqnarray}
with 
\begin{equation}
\sigma_2= -\frac{e}{c}{\bf A}\cdot {\bm \xi}.
\end{equation}
We look at the one-form $\tilde{\gamma}_1+\tilde{\gamma}_2+{\rm d}\sigma_2$ and determine the unknown functions ${\bm \xi}$, ${\cal W}_\parallel$, ${\cal W}_\perp$ and ${\cal T}$ such that this one-form no longer possesses $\Theta$-dependent terms. 
We notice that the spatial derivatives of ${\cal T}$ and ${\bm \xi}$ are of order $\epsilon_B$; therefore, the terms in ${\rm d}{\bm\xi}$ and ${\rm d}{\cal T}$ only contain terms in ${\rm d} W_\parallel$, ${\rm d} W_\perp$ and ${\rm d} \Theta$. For the same reason, the terms in ${\rm d}{\bm\rho}_0$ in $\tilde{\gamma}_1$ only involve terms in ${\rm d} W_\perp$ and ${\rm d} \Theta$. 

Combining Eqs.~(\ref{eqn:gamma1}) and (\ref{eqn:gamma2}), the one-form $\tilde{\gamma}_1+\tilde{\gamma}_2+{\rm d}\sigma_2$ can be written as 
\begin{equation}
\tilde{\gamma}_1+\tilde{\gamma}_2+{\rm d}\sigma_2= {\bm \Gamma}_{\bf X} \cdot {\rm d}{\bf X}+{\Gamma}_\Theta {\rm d}\Theta+\Gamma_{\parallel}{\rm d}W_\parallel+\Gamma_{\perp}{\rm d}W_\perp.
\end{equation}
The terms in ${\rm d}{\bf X}$ are
\begin{eqnarray}
 {\bm \Gamma}_{\bf X}&=& \frac{eB}{ c}\hat{\bf b}\times{\bm \xi}+m{\cal W}_\parallel \hat{\bf b}+\frac{m^2 c}{eB}\left[ \frac{W_\perp^2}{2B} (\hat{\bm\rho}\cdot \nabla) (B\hat{\bf b})\times \hat{\bm\rho}+W_\perp^2 (\hat{\bm\rho}\cdot\nabla)\hat{\bm\perp}-\frac{W_\perp^2 }{2}  \nabla \hat{\bm\perp}\cdot \hat{\bm\rho}\right.\nonumber\\
&&\qquad\qquad\qquad\qquad\qquad\qquad\qquad+\left.W_\parallel W_\perp (\nabla\times \hat{\bf b})\times\hat{\bm\rho}\right].
\end{eqnarray}
The terms in ${\rm d}\Theta$ are
\begin{eqnarray}
 {\Gamma}_\Theta&=& mW_\parallel \hat{\bf b}\cdot \frac{\partial {\bm \xi}}{\partial \Theta} + \frac{m^2W_\perp^2c}{2eB}\frac{\partial {\cal T}}{\partial \Theta}+ \frac{m^2c}{eB}W_\perp {\cal W}_\perp\nonumber \\
&& \qquad +\frac{m^3c^2}{e^2B^2}\left[ -\frac{W_\perp^3}{3B}\hat{\bm\rho}\cdot \nabla B+W_\parallel W_\perp^2 (\hat{\bm \rho}\cdot \nabla)\hat{\bf b} \cdot \hat{\bm\perp}\right],
\end{eqnarray}
where we have used the identities $\nabla \hat{\bf b}\cdot \hat{\bf b}=\nabla \hat{\bm\perp}\cdot \hat{\bm\perp}=0$ since $\hat{\bf b}$ and $\hat{\bm\perp}$ are of unit norm.
The terms in ${\rm d}W_\parallel$ are
\begin{equation}
 \Gamma_{\parallel}=mW_\parallel \hat{\bf b}\cdot \frac{\partial {\bm \xi}}{\partial W_\parallel}+\frac{m^2W_\perp^2 c}{2eB}\frac{\partial {\cal T}}{\partial W_\parallel},
\end{equation}
and the terms in ${\rm d}W_\perp$ are
\begin{equation}
 \Gamma_{\perp}= mW_\parallel \hat{\bf b}\cdot \frac{\partial {\bm\xi}}{\partial W_\perp}+\frac{m^2W_\perp^2 c}{2eB}\frac{\partial {\cal T}}{\partial W_\perp}+\frac{m^3 c^2}{e^2B^2}\left[W_\perp^2(\hat{\bm\rho}\cdot \nabla) \hat{\bm\perp}\cdot \hat{\bm\rho}+ W_\parallel W_\perp(\hat{\bm\rho}\cdot \nabla) \hat{\bf b}\cdot \hat{\bm\rho}\right].
\end{equation}
The terms in ${\bm\Gamma}_{\bf X}$ perpendicular to $\hat{\bf b}$ determine the perpendicular component of ${\bm\xi}$, whereas the terms parallel to $\hat{\bf b}$ determine ${\cal W}_\parallel$. The perpendicular component of ${\bm\xi}$ is obtained from the cross-product of ${\bf \Gamma}_{\bf X}$ with $\hat{\bf b}$, and is given by
\begin{equation}
 {\bm\xi}^\perp=\frac{m^2c^2}{e^2B^2}\left(-\frac{W_\perp^2}{2B}(\hat{\bm\rho}\cdot \nabla B) \hat{\bm\rho}+W_\perp^2\hat{\bf b}\times (\hat{\bm\rho}\cdot\nabla)\hat{\bm\perp} -W_\parallel W_\perp (\hat{\bf b}\cdot \nabla \times \hat{\bf b}) \hat{\bm\rho} \right),
\end{equation}
where we have used the fact that $\hat{\bf b}\times[(\hat{\bm\rho}\cdot\nabla)\hat{\bf b}\times\hat{\bm\rho}]=0$ which comes from $\nabla \hat{\bf b}\cdot \hat{\bf b}=0$.
The scalar product between $\Gamma_{\bf X}$ and $\hat{\bf b}$ leads to
\begin{equation}
 {\cal W}_\parallel=\frac{mc}{eB}\left(\frac{W_\perp^2}{2}(\hat{\bm\rho}\cdot\nabla)\hat{\bf b}\cdot\hat{\bm\perp}-W_\parallel W_\perp (\nabla\times \hat{\bf b})\cdot \hat{\bm \perp} \right),
\end{equation}
where we have used the identity $\nabla \hat{\bm\perp}\cdot \hat{\bf b}=-\nabla \hat{\bf b}\cdot \hat{\bm\perp}$. 
In ${\bm \Gamma}_{\bf X}$, which this choice of functions, it remains
\begin{equation}
 {\bm \Gamma}_{\bf X}=-\frac{mc}{e}\mu {\bf R},
\end{equation}
where ${\bf R}=\nabla \hat{\bm\perp}\cdot \hat{\bm\rho}=\nabla \hat{\bf b}_1\cdot \hat{\bf b}_2$ which is independent of the gyroangle $\Theta$. 

The first two terms in $\Gamma_\Theta$, $\Gamma_\parallel$ and $\Gamma_\perp$ calls for a gauge transformation with 
\begin{equation}
 \sigma_3=-m W_\parallel \hat{\bf b}\cdot {\bm\xi}- \frac{m^2W_\perp^2 c}{2eB}   {\cal T}.
\end{equation}
The terms in $\Gamma_\Theta$, $\Gamma_\parallel$ and $\Gamma_\perp$ becomes
\begin{eqnarray}
\widetilde{\Gamma}_\Theta &=&  \frac{m^2 c}{eB}W_\perp {\cal W}_\perp +\frac{m^3c^2}{e^2B^2}\left[-\frac{W_\perp^3}{3B}\hat{\bm\rho}\cdot \nabla B+W_\parallel W_\perp^2(\hat{\bm\rho}\cdot \nabla )\hat{\bf b}\cdot \hat{\bm\perp}\right],\\
\widetilde{\Gamma}_\parallel &=& -m\hat{\bf b}\cdot {\bm\xi},\\
\widetilde{\Gamma}_\perp &=&  - \frac{m^2 c}{eB}W_\perp{\cal T} +\frac{m^3 c^2}{e^2B^2} \left[W_\perp^2(\hat{\bm\rho}\cdot \nabla) \hat{\bm\perp}\cdot \hat{\bm\rho}+ W_\parallel W_\perp(\hat{\bm\rho}\cdot \nabla) \hat{\bf b}\cdot \hat{\bm\rho}\right].
\end{eqnarray}
Using this gauge transformation, the equations which determine the unknown functions ${\cal W}_\perp$, ${\cal W}_\parallel$ and the parallel component of ${\bm\xi}$ have been decoupled. We choose $\cal T$ such that it only eliminates the fluctuating terms. Using the following expression for the fluctuating part of $(\hat{\bm\rho}\cdot\nabla )\hat{\bf b}\cdot\hat{\bm\rho}$ :
\begin{equation}
\label{eq:rhorho}
\oversortoftilde{(\hat{\bm\rho}\cdot\nabla )\hat{\bf b}\cdot\hat{\bm\rho}}=\frac{1}{2}\left((\hat{\bm\rho}\cdot\nabla )\hat{\bf b}\cdot\hat{\bm\rho}-(\hat{\bm\perp}\cdot\nabla )\hat{\bf b}\cdot\hat{\bm\perp}\right), 
\end{equation}
we obtain the following expression for $\cal T$: 
\begin{equation}
 {\cal T} = \frac{m c}{eB} \left[W_\perp (\hat{\bm\rho}\cdot \nabla)\hat{\bm\perp}\cdot \hat{\bm\rho} + \frac{W_\parallel}{2}\left( (\hat{\bm\rho}\cdot \nabla)\hat{\bf b}\cdot \hat{\bm\rho}-(\hat{\bm\perp}\cdot \nabla)\hat{\bf b}\cdot \hat{\bm\perp}\right)\right].
\end{equation}
In the one form it remains a term in ${\rm d} W_\perp$:
\begin{equation}
 \frac{m^3 c^2}{2e^2B^2} W_\parallel W_\perp \left[ (\hat{\bm\rho}\cdot \nabla)\hat{\bf b}\cdot \hat{\bm\rho}+(\hat{\bm\perp}\cdot \nabla)\hat{\bf b}\cdot \hat{\bm\perp}\right] {\rm d}W_\perp.
\end{equation}
Using a gauge transformation with 
\begin{equation}
 \sigma_4= -\frac{m^3 c^2}{4e^2B^2} W_\parallel W_\perp^2 \left[ (\hat{\bm\rho}\cdot \nabla)\hat{\bf b}\cdot \hat{\bm\rho}+(\hat{\bm\perp}\cdot \nabla)\hat{\bf b}\cdot \hat{\bm\perp}\right],
\end{equation}
the term in ${\rm d} W_\perp$ is eliminated and a new term appears in ${\rm d}W_\parallel$, which then determines the parallel component of $\bm\xi$:
\begin{equation}
 \hat{\bf b}\cdot {\bm\xi} = -\frac{m^2c^2}{4e^2B^2}W_\perp^2\left[ (\hat{\bm\rho}\cdot \nabla)\hat{\bf b}\cdot \hat{\bm\rho}+(\hat{\bm\perp}\cdot \nabla)\hat{\bf b}\cdot \hat{\bm\perp}\right]. 
\end{equation}
For ${\cal W}_\perp$, we choose 
\begin{equation}
 {\cal W}_\perp =\frac{mc}{eB}\left[\frac{W_\perp^2}{3B}\hat{\bm\rho}\cdot \nabla B-W_\parallel W_\perp(\hat{\bm\rho}\cdot \nabla )\hat{\bf b}\cdot \hat{\bm\perp}\right],
\end{equation}
in order to eliminate terms in ${\rm d} \Theta$. 
Other choices of functions ${\bm\xi}$, ${\cal W}_\parallel$, ${\cal W}_\perp$ and $\cal T$ are possible and lead to equivalent theories, also when pushed to high orders [see \cite{Tronko_Brizard_2015,Burby_2013,Parra_2011,Parra_2014} for more details]. 
Using the change of variables defined by Eq.~(\ref{eq:cccc}) with the functions defined above, the $\Theta$-dependence has been removed from the symplectic part of the one-form. However there is still some dependence in $\Theta$ in the Hamiltonian

Now, we proceed with the last step which is a canonical Lie transform of the Hamiltonian as in Sec.~\ref{sec:Lie-transform}. This transformation does not affect the symplectic part of the one-form and is only designed to eliminate the $\Theta$-dependence in the Hamiltonian. 
With the change of coordinates (\ref{eq:cccc}), the Hamiltonian has been changed into
\begin{equation}
H=\frac{1}{2}mW_\parallel^2+\frac{1}{2}mW_\perp^2+ m(W_\parallel{\cal W}_\parallel +W_\perp {\cal W}_\perp) +{\cal O}(\epsilon_B^2). 
\end{equation}
We consider a canonical Lie transform with the following generating function
\begin{equation}
 \frac{\partial S}{\partial \Theta}+\frac{m^2c}{eB}(W_\parallel\widetilde{\cal W}_\parallel +W_\perp \widetilde{\cal W}_\perp)=0,
\end{equation}
where $\widetilde{\cal W}_\parallel$ and $\widetilde{\cal W}_\perp$ are the fluctuating part of ${\cal W}_\parallel$ and ${\cal W}_\perp$. We notice that the gradient of the generating function, i.e., $ \nabla S$, is of order $\epsilon_B^2$. The way to determine the generating function $S$ follows from the same principle as briefly explained in Sec.~\ref{sec:Lie-transform}. 
The Hamiltonian becomes
\begin{equation}
H=\frac{1}{2}mW_\parallel^2+\frac{1}{2}mW_\perp^2+ m(W_\parallel\langle{\cal W}_\parallel\rangle +W_\perp \langle {\cal W}_\perp\rangle) = \frac{1}{2}mW_\parallel^2+\mu B+  \frac{mc}{2e}W_\parallel \mu \hat{\bf b} \cdot (\nabla \times \hat{\bf b}),
\end{equation}
where $\mu=mW_\perp^2/(2B)$. 
Here we have used the identity $\langle (\hat{\bm\rho}\cdot \nabla)\hat{\bf b} \cdot \hat{\bm\perp}\rangle=-(\hat{\bf b}\cdot (\nabla\times \hat{\bf b}))/2$. In order to complete the guiding-centre derivation, we need to specify the full change of coordinates, which is a combination of the change given by Eqs.~(\ref{eq:cX})-(\ref{eq:ct}) and the canonical Lie transform, which is, up to order $\epsilon_B^2$ terms:
\begin{eqnarray}
&& \bar{\bf x}={\bf X}+ {\bm \xi}+\{S,{\bf X}\}_{\rm gc}={\bf X}+ {\bm \xi}^\perp+(\hat{\bf b}\cdot {\bm\xi})\hat{\bf b}-\frac{\hat{\bf b}}{m}\frac{\partial S}{\partial W_\parallel},\\
&& w_\parallel = W_\parallel + {\cal W}_\parallel +\{S,W_\parallel\}_{\rm gc}=W_\parallel + {\cal W}_\parallel,\\
&&  w_\perp =  W_\perp + {\cal W}_\perp +\{S,W_\perp\}_{\rm gc}=W_\perp + {\cal W}_\perp+ \frac{eB}{m^2 W_\perp c}\frac{\partial S}{\partial \Theta},\\
&& \theta = \Theta+ {\cal T} +\{S,\Theta\}_{\rm gc}=\Theta+ {\cal T}- \frac{eB}{m^2 W_\perp c}\frac{\partial S}{\partial W_\perp},
\end{eqnarray}
where the Poisson bracket is obtained from the symplectic part of the one-form, and is given by 
\begin{equation}
 \{F,G\}_{\rm gc}=\frac{e}{m c}\left( \frac{\partial F}{\partial \Theta}\frac{\partial G}{\partial \mu}-\frac{\partial F}{\partial \mu}\frac{\partial G}{\partial \Theta}\right)+\frac{{\bf B}^*}{mB_\parallel^*}\cdot \left( \nabla^* F\frac{\partial G}{\partial {W}_\parallel}-\frac{\partial F}{\partial {W}_\parallel}\nabla^* G\right)- \frac{c\hat{\bf b}}{eB_\parallel^*}\cdot (\nabla^* F\times \nabla^* G),
\end{equation}
where 
\begin{eqnarray}
&& \nabla^*=\nabla-\mathbf{R}\ \frac{\partial}{\partial\theta},\\
&&  {\bf B}^*={\bf B}+\frac{mc}{e} W_\parallel \nabla\times\hat{\bf b}-\frac{mc^2}{e^2}\mu \nabla\times {\bf R},  
\end{eqnarray}
and $B_\parallel^*={\bf B}^*\cdot \hat{\bf b}$. In order to obtain explicit expressions for the change of coordinates, we specify the fluctuating part of the functions ${\cal W}_\parallel$ and ${\cal W}_\perp$:
\begin{eqnarray}
  \widetilde{\cal W}_\perp &=&\frac{mc}{eB}\left( \frac{W_\perp^2}{3B}\hat{\bm\rho}\cdot\nabla B-\frac{W_\parallel W_\perp}{2}\left((\hat{\bm\rho}\cdot\nabla)\hat{\bf b}\cdot \hat{\bm\perp}+(\hat{\bm\perp}\cdot\nabla)\hat{\bf b}\cdot \hat{\bm\rho} \right)\right),\\
 \widetilde{\cal W}_\parallel &=& \frac{mc}{eB}\left( \frac{W_\perp^2}{4}\left((\hat{\bm\rho}\cdot\nabla)\hat{\bf b}\cdot \hat{\bm\perp}+(\hat{\bm\perp}\cdot\nabla)\hat{\bf b}\cdot \hat{\bm\rho} \right) -W_\parallel W_\perp (\nabla \times \hat{\bf b})\cdot \hat{\bm\perp}\right),
\end{eqnarray}
where we have used the fact that the fluctuating part of $(\hat{\bm\rho}\cdot\nabla)\hat{\bf b}\cdot \hat{\bm\perp}$ is given by
\begin{equation}
\oversortoftilde{(\hat{\bm\rho}\cdot\nabla)\hat{\bf b}\cdot \hat{\bm\perp}}=\frac{1}{2}\left((\hat{\bm\rho}\cdot\nabla)\hat{\bf b}\cdot \hat{\bm\perp}+(\hat{\bm\perp}\cdot\nabla)\hat{\bf b}\cdot \hat{\bm\rho} \right). 
\end{equation}
The generating function $S$ is chosen to be purely fluctuating: 
\begin{equation}
  S = \frac{m^3c^2}{e^2B^2}\left[\frac{W_\perp^3}{3B} \hat{\bm\perp}\cdot \nabla B+ \frac{W_\parallel W_\perp^2}{8} \left((\hat{\bm\rho}\cdot\nabla )\hat{\bf b}\cdot\hat{\bm\rho}-(\hat{\bm\perp}\cdot\nabla )\hat{\bf b}\cdot\hat{\bm\perp}\right)+W_\parallel^2 W_\perp (\nabla \times \hat{\bf b})\cdot \hat{\bm\rho} \right],
\end{equation}
where we have used the fluctuating part of $(\hat{\bm\rho}\cdot\nabla )\hat{\bf b}\cdot\hat{\bm\rho}$ is given by Eq.~(\ref{eq:rhorho}).
Therefore the change of coordinates from $({\bf x},{\bf v})$ to $({\bf X},W_\parallel,W_\perp,\Theta)$ is given by
\label{eq:chgt}
\begin{eqnarray}
&&  {\bf x} = {\bf X}+\frac{m W_\perp c}{eB}\hat{\bm\rho}+\frac{m^2c^2}{e^2B^2}\left(-\frac{W_\perp^2}{2B}\left( (\hat{\bm\rho}\cdot \nabla B) \hat{\bm\rho}+2(\hat{\bm\perp}\cdot \nabla B) \hat{\bm\perp}\right) \right.\nonumber \\
&&\qquad\qquad\qquad\left.-W_\parallel W_\perp \left( (\hat{\bf b}\cdot  \nabla \times \hat{\bf b}) \hat{\bm\rho}+2(\hat{\bm\rho}\cdot  \nabla \times \hat{\bf b}) \hat{\bf b} \right)\right. \nonumber \\
&& \qquad\qquad\qquad  -\frac{W_\perp^2}{8}\left(3\hat{\bm\rho}\cdot \nabla\hat{\bf b}\cdot \hat{\bm\rho}+\hat{\bm\perp}\cdot \nabla\hat{\bf b}\cdot \hat{\bm\perp} \right)\hat{\bf b}+W_\parallel^2 \left((\hat{\bm\perp}\cdot\nabla \times \hat{\bf b} )\hat{\bm\rho}-(\hat{\bm\rho}\cdot\nabla \times \hat{\bf b})\hat{\bm\perp} \right)\nonumber \\
&& \qquad\qquad\qquad   \left. +\frac{W_\parallel W_\perp}{4} \left(-3 \hat{\bm \rho}\cdot \nabla\hat{\bf b}\cdot\hat{\bm\perp}+ \hat{\bm \perp}\cdot \nabla\hat{\bf b}\cdot\hat{\bm\rho} \right) \hat{\bm\rho} \right. \nonumber\\
&&\left. \qquad\qquad\qquad+\frac{W_\parallel W_\perp}{4}\left( \hat{\bm \rho}\cdot \nabla\hat{\bf b}\cdot\hat{\bm\rho}-\hat{\bm \perp}\cdot \nabla\hat{\bf b}\cdot\hat{\bm\perp} \right)\hat{\bm\perp} \right),\label{eq:chgt1}\\
&& v_\parallel = W_\parallel+\frac{mc}{eB}\left(\frac{W_\perp^2}{2}\hat{\bm\rho}\cdot\nabla\hat{\bf b}\cdot\hat{\bm\perp}-W_\parallel W_\perp \nabla\times \hat{\bf b}\cdot \hat{\bm \perp} \right),\\
&&  v_\perp = W_\perp +\frac{mc}{eB}\left(-\frac{3W_\parallel W_\perp}{4} \hat{\bm\rho}\cdot \nabla\hat{\bf b}\cdot \hat{\bm\perp}+\frac{W_\parallel W_\perp}{4}\hat{\bm\perp}\cdot\nabla\hat{\bf b}\cdot \hat{\bm\rho} +W_\parallel^2 \nabla\times \hat{\bf b}\cdot \hat{\bm\perp} \right),\\
&&  \theta = \Theta + \frac{m c}{eB} \left(-\frac{W_\perp}{B}\hat{\bm\perp}\cdot \nabla B+W_\perp \hat{\bm\rho}\cdot \nabla\hat{\bm\perp}\cdot \hat{\bm\rho}+ \frac{W_\parallel}{4}\left( \hat{\bm\rho}\cdot \nabla\hat{\bf b}\cdot \hat{\bm\rho}-\hat{\bm\perp}\cdot \nabla\hat{\bf b}\cdot \hat{\bm\perp}\right)
\right.\nonumber \\
&&\qquad\qquad \qquad\qquad\qquad \left. -\frac{W_\parallel^2}{W_\perp}\nabla \times \hat{\bf b}\cdot \hat{\bm\rho}\right),\label{eq:chgt2}
\end{eqnarray}
where the right hand side is evaluated at $({\bf X},W_\parallel,W_\perp,\Theta)$ and where
\begin{equation}
{\bf v}=v_\parallel \hat{\bf b}({\bf x})+v_\perp \hat{\bm\perp}(\theta,{\bf x}).
\end{equation}
The inversion of the change of variables given by Eq.~(\ref{eq:chgt}), i.e., providing $({\bf X},W_\parallel,W_\perp,\Theta)$ as functions of $({\bf x},v_\parallel,v_\perp,\theta)$, gives the exact same equations given in \cite{Littlejohn_1983}.

We note that the term $W_\parallel \mu \hat{\bf b} \cdot (\nabla \times \hat{\bf b})$ is usually moved to the symplectic part of the one-form by a translation in $W_\parallel$. In the new coordinates, it leads to the Hamiltonian
\begin{equation}
H_{\rm gc}=\mu B({\bf X})+\frac{1}{2}mW_\parallel^2,
\end{equation}
and the one-form
\begin{equation}
 \gamma_{\rm gc}=\left[\frac{e}{c}{\bf A}({\bf X})+mW_\parallel \hat{\bf b}({\bf X})- \frac{mc}{e}\mu {\bf R}^* \right]\cdot {\rm d}{\bf X}+ \frac{mc}{e}\mu {\rm d}\Theta-H_{\rm gc} {\rm d}t,
\end{equation}
where ${\bf R}^*=\nabla \hat{\bf b}_1\cdot \hat{\bf b}_2+(\hat{\bf b}\cdot \nabla\times\hat{\bf b})\hat{\bf b}/2$. 


\begin{thebibliography}{10}

\bibitem{Garbet_Idomura_2010}
X.~{Garbet}, Y.~{Idomura}, L.~{Villard}, and T.~H. {Watanabe}.
\newblock {Gyrokinetic simulations of turbulent transport}.
\newblock {\em {Nuclear Fusion}}, 50:043002, 2010.

\bibitem{Schekochihin_2009}
A.~A. Schekochihin, S.~C. Cowley, W.~Dorland, G.~W. Hammett, G.~G. Howes,
  E.~Quataert, and T.~Tatsuno.
\newblock Astrophysical gyrokinetics: kinetic and fluid turbulent cascades in
  magnetized weakly collisional plasmas.
\newblock {\em Astrophysical Journal Supplement}, 182:310, 2009.

\bibitem{Brizard_Hahm}
A.~J. {Brizard} and T.~S. {Hahm}.
\newblock {Foundations of nonlinear gyrokinetic theory}.
\newblock {\em Reviews of Modern Physics}, 79:421, 2007.

\bibitem{Jolliet_2007}
S.~{Jolliet}, A.~{Bottino}, P.~{Angelino}, R.~{Hatzky}, T.~M. {Tran}, B.~F.
  {Mcmillan}, O.~{Sauter}, K.~{Appert}, Y.~{Idomura}, and L.~{Villard}.
\newblock {A global collisionless PIC code in magnetic coordinates}.
\newblock {\em Computer Physics Communications}, 177:409, 2007.

\bibitem{Peeters_2009}
A.~G. Peeters, Y.~Camenen, F.~J. Casson, W.~A. Hornsby, A.~P. Snodin,
  D.~Strintzi, and G.~Szepesi.
\newblock The nonlinear gyro-kinetic flux tube code gkw.
\newblock {\em Computer physics communications}, 180(12):2650--2672, 2009.

\bibitem{Goerler_2011}
T.~{Goerler}, X.~{Lapillonne}, S.~{Brunner}, {Dannert} T., F.~{Jenko},
  F.~{Merz}, and D.~{Told}.
\newblock {The global version of the gyrokinetic turbulence code {GENE} }.
\newblock {\em {Journal of Computational Physics}}, 230:7053 -- 7071, 2011.

\bibitem{Grandgirard_2016}
V.~Grandgirard, J.~Abiteboul, J.~Bigot, J.~Cartier-Michaud, N.~Crouseillese,
  G.~Dif-Pradalier, Ch. Ehrlacher, D.~Esteve, X.~Garbet, Ph. Ghendrih, G.~Latu,
  M.~Mehrenberger, C.~Norscini, Ch. Passeron, F.~Rozar, Y.~Sarazin,
  E.~Sonnendruecker, A.~Strugarek, and D.~Zarzoso.
\newblock A 5d gyrokinetic full-f global semi-lagrangian code for flux-driven
  ion turbulence simulations.
\newblock {\em Computer physics communications}, 207:35--68, 2016.

\bibitem{Candy_2003}
J.~Candy and R.~E. Waltz.
\newblock An eulearian gyrokinetic-maxwell solver.
\newblock {\em Journal of Computational Physics}, 186:545, 2003.

\bibitem{Casati_2009}
A~Casati, T~Gerbaud, P~Hennequin, C~Bourdelle, J~Candy, F~Clairet, X~Garbet,
  V~Grandgirard, O~D Gürcan, S~Heuraux, G~T Hoang, C~Honoré, F~Imbeaux,
  R~Sabot, Y~Sarazin, L~Vermare, and R~E Waltz.
\newblock Turbulence in the tore supra tokamak: measurements and validation of
  nonlinear simulations.
\newblock {\em Physical review letters}, 102(16):165005--165005, 2009.

\bibitem{Wersal_2012}
C.~Wersal, A.~Bottino, P.~Angelino, and B.~D. Scott.
\newblock Fluid moments and spectral diagnostics in global particle-in-cell
  simulations.
\newblock {\em Journal of physics. Conference series}, 401, 2012.

\bibitem{CATTO1978}
P.~J. Catto.
\newblock Linearized gyro-kinetics.
\newblock {\em Plasma Physics and Controlled Fusion}, 20(7):719--722, 1978.

\bibitem{CATTO1981}
P.~J. Catto.
\newblock Generalized gytokonetics.
\newblock {\em Plasma Physics and Controlled Fusion}, 23(7):639--650, 1981.

\bibitem{Frieman_Chen_1982}
E.~A. {Frieman} and L.~{Chen}.
\newblock {Nonlinear gyrokinetic equations for low-frequency electromagnetic
  waves in general plasma equilibria}.
\newblock {\em Physics of Fluids}, 25:502, 1982.

\bibitem{Abel_2013}
I.~G. Abel, G.~G. Plunk, E.~Wang, M.~Barnes, S.~C. Cowley, W.~Dorland, and
  A.~A. Schekochihin.
\newblock Multiscale gyrokinetics for rotating tokamak plasmas: fluctuations,
  transport and energy flows.
\newblock {\em Report on Progress in Physics}, 76:116201, 2013.

\bibitem{Littlejohn_1979}
R.~G. {Littlejohn}.
\newblock {A guiding center {H}amiltonian: a new approach.}
\newblock {\em {Journal of Mathematical Physics}}, 20:2445, 1979.

\bibitem{Littlejohn_1981}
R.~G. {Littlejohn}.
\newblock {Hamiltonian formulation of guiding center motion}.
\newblock {\em {Physics of Fluids}}, 29:1730, 1981.

\bibitem{Littlejohn_1983}
R.~G. {Littlejohn}.
\newblock {Variational principles of guiding centre motion}.
\newblock {\em {Journal of Plasma Physics}}, 29:111, 1983.

\bibitem{Sugama_2000}
H.~{Sugama}.
\newblock {Gyrokinetic field theory}.
\newblock {\em Physics of Plasmas}, 7:466, 2000.

\bibitem{brizard_prl_2000}
A.~J. {Brizard}.
\newblock {New Variational Principle for the {V}lasov{-}{M}axwell Equations}.
\newblock {\em Physical Review Letters}, 84:5768, 2000.

\bibitem{Squire_Qin_2013}
J.~{Squire}, H.~{Qin}, W.~M. {Tang}, and C.~{Chandre}.
\newblock {The Hamiltonian structure and {E}uler{-}{P}{oincar\' e} formulation
  of the {V}lasov{-}{M}axwell and gyrokinetic systems}.
\newblock {\em {Physics of Plasmas}}, 20:022501, 2013.

\bibitem{Burby_2015}
J.~W. Burby, A.~J. Brizard, P.~J. Morrison, and H.~Qin.
\newblock Hamiltonian gyrokinetic {V}lasov {-} {M}axwell system.
\newblock {\em Physics Letters A}, 379:2073, 2015.

\bibitem{TBS_2016}
N.~{Tronko}, A.~{Bottino}, and E.~{Sonnendr\"ucker}.
\newblock {Second order gyrokinetic theory for {P}article{-}{I}n{-}{C}ell
  codes}.
\newblock {\em {Physics of Plasmas}}, 23:082505, 2016.

\bibitem{PPCF_2017}
N.~{Tronko}, A.~{Bottino}, C.~{Chandre}, and E.~{Sonnendr\"ucker}.
\newblock {Hierarchy of second order gyrokinetic {H}amiltonian models for
  {P}article{-}{I}n{-}{C}ell codes}.
\newblock {\em {Plasma Physics and Controlled Fusion}}, 59:064008, 2017.

\bibitem{Hahm_1988}
T.~S. {Hahm}.
\newblock {Nonlinear gyrokinetic equations for tokamak microturbulence}.
\newblock {\em Physics of Fluids}, 31:2670, 1988.

\bibitem{Brizard_1989}
A.~J. {Brizard}.
\newblock {Nonlinear gyrokinetic {M}axwell{-}{V}lasov equations using magnetic
  co{-}ordinates}.
\newblock {\em Journal of Plasma Physics}, 41:541, 1989.

\bibitem{Grebogi_1979}
C.~Grebogi, A.~N. Kaufman, and R.~G. Littlejohn.
\newblock Hamiltonian theory of ponderomotive effects of an electromagnetic
  wave in a nonuniform magnetic field.
\newblock {\em Physical Review Letters}, 43:22, 1979.

\bibitem{Cary_Brizard}
J.~R. {Cary} and A.~J. {Brizard}.
\newblock {Hamiltonian theory of guiding center motion}.
\newblock {\em Reviews of Modern Physics}, 81:693, 2009.

\bibitem{Tronko_Brizard_2015}
N.~{Tronko} and A.~J. {Brizard}.
\newblock {{L}agrangian and {H}amiltonian constraints for guiding center
  {H}amiltonian theories}.
\newblock {\em Physics of Plasmas}, 22:112507, 2015.

\bibitem{Cary_1981}
J.~R. Cary.
\newblock {Lie transform perturbation theory for Hamiltonian systems}.
\newblock {\em {Physics Reports}}, 79:129, 1981.

\bibitem{Bottino_Sonnendruecker}
A.~{Bottino} and E.~{Sonnendr\"ucker}.
\newblock {Monte Carlo Particle-In-Cell methods for the simulation of the
  Vlasov-Maxwell gyrokinetic equations}.
\newblock {\em Journal of Plasma Physics}, 81:435810501, 2015.

\bibitem{Krause_2007}
T.~B. Krause, A.~Apte, and P.~J. Morrison.
\newblock A unified approach to the darwin approximation.
\newblock {\em Physics of Plasmas}, 14:102112, 2007.

\bibitem{Burby_2013}
J.~W. Burby, J.~Squire, and H~Qin.
\newblock Automation of the guiding center expansion.
\newblock {\em Physics of Plasmas}, 20:072105, 2013.

\bibitem{Parra_2011}
F.~I. Parra and I~Calvo.
\newblock Phase-space {L}agrangian derivation of electrostatic gyrokinetics in
  general geometry.
\newblock {\em Plasma Physics and Controlled Fusion}, 53:045001, 2011.

\bibitem{Parra_2014}
F.~I. Parra, I~Calvo, J.~W. Burby, and H.~Qin.
\newblock Equivalence of two independent calculations of the higher order
  guiding center {L}agrangian.
\newblock {\em Physics of Plasmas}, 21:104506, 2014.

\end{thebibliography}

\end{document}